\documentclass[review]{cas-sc}

\usepackage[numbers]{natbib}

\usepackage{longtable} 
\usepackage{changepage} 
\usepackage{multirow}
\usepackage{xcolor}
\usepackage{pifont} 
 \usepackage{kantlipsum} 

\def\tsc#1{\csdef{#1}{\textsc{\lowercase{#1}}\xspace}}
\tsc{WGM}
\tsc{QE}
\tsc{EP}
\tsc{PMS}
\tsc{BEC}
\tsc{DE}

\newcommand{\sectopic}[1]{\vspace{0.2em}\par\noindent{\textit{\bfseries #1}}}
\newcommand{\ca}[1]{\todo[color=blue!40]{CA: #1}}

\begin{document}
\let\WriteBookmarks\relax
\def\floatpagepagefraction{1}
\def\textpagefraction{.001}

\shorttitle{Requirements Practices and Gaps for Human-centered AI}

\shortauthors{Ahmad et~al.}

\title [mode = title]{Requirements Practices and Gaps When Engineering Human-Centered Artificial Intelligence Systems}                      



\author[1]{Khlood Ahmad}


\ead{ahmadkhl@deakin.edu.au}

\author[1]{Mohamed Abdelrazek}
\ead{mohamed.abdelrazek@deakin.edu.au}

\author[3]{Chetan Arora}
\ead{chetan.arora@monash.edu}

\affiliation[1]{organization={Deakin University}, 
    city={Geelong},
    state={VIC},
    country={Australia}}

\author[2]{Muneera Bano}
\ead{muneera.bano@csiro.au}

\affiliation[2]{organization={CSIRO's Data61}, 
    city={Clayton},
    state={VIC},
    country={Australia}}

\author[3]{John Grundy}
\ead{john.grundy@monash.edu}

\affiliation[3]{organization={Monash University}, 
    city={Clayton},
    state={VIC},
    country={Australia}}

\begin{abstract}
\noindent [Context] Engineering Artificial Intelligence (AI) software is a relatively new area with many challenges, unknowns, and limited proven best practices. Big companies such as Google, Microsoft, and Apple have provided a suite of recent guidelines to assist engineering teams in building human-centered AI systems. [Objective] The practices currently adopted by practitioners for developing such systems, especially during Requirements Engineering (RE), are little studied and reported to date. [Method] This paper presents the results of a survey conducted to understand current industry practices in RE for AI (RE4AI) and to determine which key human-centered AI guidelines should be followed. Our survey is based on mapping existing industrial guidelines, best practices, and efforts in the literature. [Results] We surveyed 29 professionals and found most participants agreed that all the human-centered aspects we mapped should be addressed in RE. Further, we found that most participants were using UML or Microsoft Office to present requirements. [Conclusion] We identify that most of the tools currently used are not equipped to manage AI-based software, and the use of UML and Office may pose issues to the quality of requirements captured for AI. Also, all human-centered practices mapped from the guidelines should be included in RE.
\end{abstract}


\maketitle

\section{Introduction}~\label{sec:introduction}

The use of AI has become a major focus of current-day technologies, with many successes in the industry. For example, Mercedes Benz replaced standard robots with AI-powered cobots, which allowed them to manufacture customised cars with higher efficiency~\cite{wilson2018collaborative}. And the IBM Watson system could recommend cancer treatments that are $\approx$99\% of the time in line with the physician's recommendations~\cite{jiang2017artificial}. On the other hand, AI has had many failures and well-publicised biases. Examples include the Facebook chatbot that could respond correctly to only 30\% of its Messenger services~\cite{FacebookChatbot}, and Microsoft's AI chatbot that `learned' racist slurs within a day of reading Twitter feeds~\cite{price2016microsoft}. Although these systems might have addressed all functional requirements and technical goals, the outcome did not necessarily reflect the users' \emph{human-centered needs}~\cite{maguire2001methods}. For example, a technical goal to build a \emph{fast} face recognition system could be easily achievable. However, the resulting system might still discriminate against users' color or race. These {human-centered aspects} should be addressed along with the technical goals~\cite{whittle2019your}.

Schmidt~\cite{schmidt2020interactive} compares AI and ML algorithms to raw materials (such as metals in the ground). To work with these materials efficiently, we need to find the correct tools to use ``without compromising human values''. In the context of AI, human-centered approaches include, among others, providing better user experience~\cite{amershi2014power}, improved explainability~\cite{sokol2020one, dodge2019explaining, miller2017explainable}, fairness~\cite{bellamy2018ai}, trust\cite{amershi2014power}, reduced biases~\cite{roselli2019managing,amershi2014power}, and building responsible AI~\cite{dignum2017responsible}. However, today's AI software lack these human-centered aspects~\cite{amershi2014power, schmidt2020interactive}, and there is a need to research appropriate AI solutions before including them in software systems. 

The main focus of AI in recent years has been to build technically sound systems. However, there is a recent trend towards exploring more human-centered aspects when building AI software~\cite{schmidt2020interactive}. Industries such as Apple~\cite{Apple2020}, Google~\cite{GooglePair2019}, and Microsoft~\cite{Microsoft2022} are adopting (and recommending to adopt) human-centered approaches when building AI software. However, our systematic literature review (SLR)~\cite{ahmad2021SLR, ahmad2022mapping} found that most software systems lack human-centered approaches when writing and modeling requirements. The SLR showed that existing RE for AI (RE4AI) research primarily focused on certain aspects to include ethics~\cite{kuwajima2019adapting, aydemir2018roadmap}, trust~\cite{amaral2020ontology} and explainability~\cite{hall2019systematic, schoonderwoerd2021human, cirqueira2020scenario, kohl2019explainability} with limited or no empirical evaluations.  None of these existing studies investigated the human-centered aspects adopted by the industry or how they should be addressed in RE.

In this paper, we conduct a study to identify the gaps existing in current RE4AI practices. We aim to answer two research questions (RQs): \textcolor{black}{The first RQ (RQ1) focuses on mapping the existing RE4AI research and industry guidelines for developing human-centered AI and highlighting the missing human-centered aspects in RE4AI. The second RQ (RQ2) identifies the gaps between RE4AI research and industrial practices. In RQ1, we analyze several current industry guidelines for building AI software from Google~\cite{GooglePair2019}, Microsoft~\cite{Microsoft2022}, Apple~\cite{Apple2020}, and the Machine Learning (ML) Canvas~\cite{MLCanvas} (discussed in Sections~\ref{sec:Background} and \ref{sec:Checklist}), and map them against studies from our SLR on RE4AI~\cite{ahmad2021SLR}. Based on the outcomes of this mapping, we survey in RQ2 researchers and industry practitioners to determine the guidelines that need to be included in RE.}  From our survey results, we were able to find additional modeling languages, requirements notations, tools, and issues and challenges that have not yet been reported. Furthermore, we determined which key human-centered approaches should be included~in~RE4AI.

The key contributions of this research include the following:

\begin{itemize}
\item We combine human-centered guidelines from the literature and industrial guidelines and map them against six different human-centered aspects, mainly based on Google's PAIR guidelines~\cite{GooglePair2019}.
\item We conduct a practitioner survey and find which of these human-centered guidelines should be included in RE. Our survey results show that all the human-centered approaches need to be addressed during RE when building AI~software.
\item  We identify an additional 15 tools, showcase the different notations and platforms used in practice and find that most tools and methods used need to be revised to manage RE4AI.
\item We provide a comparison between RE4AI in research and practice and identify the gaps in both RE4AI and engineering human-centered AI~research. 
\item We provide five future research recommendations.
\end{itemize}
\vspace{-0.3ex}
The rest of the paper is organized as following:  Section~\ref{sec:Background}
provides a brief background on human-centered RE and human-centered AI. Section~\ref{sec:Checklist} presents the checklist for human-centered AI. Section~\ref{sec:Survey} reports the survey results up to date. Section~\ref{sec:Threats} addresses threats to validity. Section~\ref{sec:Discussion} presents gaps between literature, practice, and human-centered approaches in RE4AI, and Section~\ref{sec:Conclusion} concludes. 

\section{Background and Related Work}~\label{sec:Background}
\vspace{-2em}

 \textcolor{black}{As discussed in Section~\ref{sec:introduction}, the recent growth in AI-based software systems has created a gap in existing RE techniques. The current RE methods and tools need augmentation and rethinking for RE4AI~\cite{kondermann2013ground, agarwal2014expert, ahmad2021SLR}. Our focus in this paper is on RE practices for human-centered aspects of AI software systems. In this section, we provide background on human-centered AI and the different industry guidelines for developing human-centered AI systems. We further cover related work on RE4AI, in general, as there is not much literature on RE for human-centered AI. 
 }

\subsection{Human-centred Development of AI Software Systems}

\textcolor{black}{ 
Human-centered design of AI software systems intends to consider human needs and values as first-class citizens when building software systems~\cite{maguire2001methods,shneiderman2022human}. AI research and development practices have focused primarily focused on the technical aspects~\cite{ehsan2020human}.
Focusing only on the technical aspects and largely ignoring the human-centered aspects when building AI software can lead to severe consequences, including physical and mental harm \cite{xu2022transitioning}. These issues are further exacerbated by the blackbox nature of AI systems, which are not well understood and explained to the end users.
Xu~\cite{xu2022transitioning} compared the current AI wave to the rise of computers in the 1980s when the focus was on the technical aspects only, and many failures were emerging due to the missing human-centered aspects. However, with the growing understanding of AI software systems development, more practices are moving towards including human-centered factors and integrating ethics when building AI software.
Riedl~\cite{riedl2019human} defines AI software as ``the study and design of algorithms that perform tasks or behaviors that a person could reasonably deem to require intelligence if a human were to do it''. 
Humans, however, base their decisions on several factors, including, among others, common sense and socio-cultural beliefs. Thus, if we build intelligent systems that understand these socio-cultural beliefs, we will have AI software that is less prone to making mistakes and provide predictions that align with human behavior~\cite{riedl2019human}. Human-centered AI focuses on amplifying and augmenting human abilities and behavior, rather than displacing them~\cite{shneiderman2022human}. 
}

Schmidt~\cite{schmidt2020interactive} defines interactive human-centered AI as systems that have the ability to provide benefits to humans, are transparent, explain the risks and benefits, and allow the user to be in control of the system. However, the author explains that it is more common for developers to focus on the technical aspects of the AI component in current AI systems and often downplay or ignore many human aspects. Wang et al. \cite{wang2019designing} explain that a lack of human-centered approaches can result in biased AI. We have seen AI systems become biased on many occasions towards people from another race, gender, color, religious views, health status, people with disability, etc. \cite{hajian2016algorithmic, khomh2018software, whittaker2019disability}. Some studies have focused on identifying these issues, such as "AI Fairness 360" \cite{bellamy2019Fairness360} and `Fairtest" \cite{tramer2017fairtest}. Other studies focused on identifying toolsets used by software engineers when building AI software. A survey found that organizations were using tools such as Azure ML Studio \cite{Azure} and Amazon AWS \cite{AWS} to manage data in ML software. However, these tools may lack in domain knowledge, modeling of data analytic aspects when building ML software, and collaboration between different domain experts working on the same project \cite{khalajzadeh2018survey}. 

Several organizations have developed guidelines to target human-centered AI design and development. We cover some of the available guidelines in this section below.

\subsection{Google People+AI Research (PAIR) Guidelines}
This guidebook includes six chapters to guide developers in providing diverse AI software that adds value to the needs of users ~\cite{GooglePair2019}. The chapters include: Identifying user Needs and determining if AI is the solution needed, collecting and evaluating the data used in AI, adjusting the mental models of the AI system to match the users needs and expectations, providing explainable and trustworthy systems, designing and managing feedback and user control, and accounting for errors and managing failures. Each chapter targets an aspect of human-centered approaches when developing AI systems. The guide starts with identifying the problem and finding if AI can solve this problem and why it's needed. Is AI required to provide predictions, personalize, or make recommendations? Or is there a need for speech and language understanding, image recognition, or fraud detection?

The second chapter discusses the data involved in ML model training and the importance of complying with data requirements when collecting data by maintaining privacy and safety measures and avoiding biases by making sure that the data used is inclusive. The third chapter explains mental models and the importance of providing user explanations that allow them to create accurate mental maps of the product. Chapter Four provided a detailed description of providing explanations when building AI systems. The guide highlights the importance of providing realistic expectations of the AI model so that people don't overtrust it and only explain what is needed.

The fifth chapter covers the importance of feedback and the types of feedback available in improving AI systems. These include explicit and implicit feedback.   Finally, when designing software systems, it's always important to allow users to take control whenever needed, whether it's their preference or the system has failed to perform its functionality. The last chapter addresses errors \& failure. The formed mental model of the AI system plays a factor in how users perceive errors. How the user perceives errors might change with time, so when they are using the system for the first time they might not consider wrong recommendations as errors but after a year of using the product that perception might change. The guidelines aim to mitigate biases, provide inclusive design to AI software, and set the expectations of the AI software's limitations and capabilities. 

\subsection{Apple's human-interface guidelines}

Apple's human-interface guidelines \cite{Apple2020} include a set of guidelines for ML developers to build human-centered ML software and prioritize the users experience. The first section explains the machine learning roles that should be taken into consideration. What types of data should be considered for the system, public or private? Do we design the system to be visible or invisible? The drawback to using invisible features is that they are harder to receive feedback and cannot be explained very well. The second section looks at the types of input to use in an ML model. These include:  Explicit, implicit, calibration, and correction. Explicit feedback is provided by users when requested by the app or system and should only be asked when required. In each of these input types, the guide always stresses the importance of securing user's privacy and providing explanations as to how the user's information is shared between apps to avoid~mistrust.

The last section addresses output to include:  Mistakes, providing multiple options, and displaying confidence.   When dealing with mistakes, it is important to mitigate the effects of mistakes to avoid mistrust and minimize them with proactive features. When outputting confidence, it was important to understand what it meant before deciding how to present it. Presenting confidence also plays a part in providing trust to the user. Confidence should be provided in terms users understand – "because you" instead of "97\%" and avoid showing results when confidence is low, especially in proactive systems. Displaying confidence should fit the context. For example, you would display confidence in the form of intervals or percentages in situations where recommendations are given for statistical or numerical data.

\subsection{Microsoft's guidelines for human-AI interaction}

Amershi et al. \cite{amershi2019guidelines} at Microsoft research \cite{Microsoft2022} examined over 20 years of guidelines for human-computer interaction research and established 18 guidelines for human-AI interaction. The guidelines were proposed over four phases: Before starting a project, while the user interacts with the software, when the system does not work as expected, and monitoring the system over time.   These guidelines aim to mitigate biases, address user needs, and provide more explainable systems. The first stage included explaining to the user the system's limitations and capabilities before interacting with the system. The second stage offers a guide to what should be delivered to the user while interacting with the AI system. This includes providing context based on the users time, task, and environment while making sure to address any biases and address social norms. The third stage address how issues are handled by allowing the user to dismiss, fix and get an explanation as to why an error happened. The last stage looks at how to adjust the AI system over~time.  

\subsection{Machine Learning Canvas}

Machine Learning Canvas \cite{MLCanvas} is a framework designed to allow software engineers, data scientists, and ML specialists to work together simultaneously on a project. The canvas is used at the start of building a project when ML is used and is divided into four blocks. The first block aids in identifying what the system is used for, why it is needed, and who the users are. The second block supports data collection and how the model will learn from data. The third block focuses on predictions and how decisions are made, and the final block is around evaluation methods and matrices used. The Machine Learning Canvas does not focus on human-centered approaches. However, it supports the integration of the different ML components and focuses on all the aspects needed to start building ML software. This allowed us to see how human-centered approaches could become integrated into any available ML model and specifically during RE.

\subsection{Related Work on RE4AI}~\label{sec:IssuesInRE4AI}
\textcolor{black}{
For RE4AI literature, we considered two main sources -- (i) a SLR~\cite{ahmad2021SLR} that we conducted to analyze papers published on RE4AI from 2010-2020, and (ii) a mapping study \cite{villamizar2021requirements} that provides insights into the methods used in RE when building AI systems. We further extended our SLR with studies from 2020 to 2021 (the paper is currently under review) and extracted additional studies on RE4AI. We identified limitations (or outstanding challenges) in the existing literature on RE4AI (presented next).
We also categorized the studies as human-centered vs. technical.}
Three studies had a focus on human-centered approaches and included:
Bruno et al.~\cite{bruno2013functional} investigated the requirements for creating a human-centered social robot for the elderly and targeted emotion;  Sandkuhl~\cite{sandkuhl2019putting} emphasized the importance of understating expectations and limitations of the AI system before deciding to use AI; and Fagbola and Thakur~\cite{fagbola2019towards} list several tools that can identify and mitigate any human-centered issues related to fairness and biases. The rest of these studies were linked to ethics~\cite{kuwajima2019adapting, aydemir2018roadmap}, explainability~\cite{hall2019systematic, schoonderwoerd2021human, cirqueira2020scenario, kohl2019explainability}, and trust~\cite{amaral2020ontology}.

\textcolor{black}{\sectopic{Outstanding Challenges in RE4AI Literature.}
When analyzing the RE4AI literature, we found a number of recurring issues, listed here:
\begin{enumerate}
    \item \textbf{Overestimating and overtrusting the capabilities of AI  solutions.} Many organizations adopt AI without having the suitable expertise or the proper data format ~\cite{sandkuhl2019putting, dimatteo2020requirements}.    
    \item \textbf{AI requirements are hard to specify.}  For example, defining a pedestrian for a self-driving car or defining ethical and explainability requirements.~\cite{cysneiros2018software,rahimi2019toward, balasubramaniam2022transparency}.
    \item \textbf{AI requirements are vague or very high-level.} Given the black box nature of most AI models, requirements engineers find it difficult to specify precisely the requirements of such systems and end up specifying requirements that are deemed `too high-level' or `vague'~\cite{martinez2022software,lu2022software}.
    \item \textbf{Limitations of existing RE techniques to manage AI requirements}, as most techniques and tools are not equipped to handle AI software~\cite{vogelsang2019requirements, nakamichi2020requirements}.
    \item \textbf{Capturing and specifying the trade-offs that might arise when building AI systems.} For example, how do we calculate the trade-off of choosing between precision vs recall, or how do you specify the trade-off between privacy, security, and explainability~\cite{dimatteo2020requirements, berry2022requirements}? 
    \item	\textbf{The  emergence  of  new types of requirements}, such as data, explainability, transparency, compliance and ethics~\cite{vogelsang2019requirements, bosch2018takes,abualhaija2022automated}.    
    \item \textbf{Issues with data requirements}, such as lack of structure, availability or quality~\cite{challa2020faulty, shin2019data,weihrauch2018conceptual, nakamichi2020requirements,altarturi2017requirement, ries2021mde}.    
    \item \textbf{The difficulties in understanding and specifying non-functional requirements in AI systems.} Certain NFR categories in AI systems, such as fairness and transparency, hold more importance than other categories, such as modularity in the traditional systems~\cite{horkoff2019nonFunctional,habibullah2021non, kohl2019explainability,cysneiros2020non}.
    \item \textbf{Difficult to understand the feasibility of AI models and the outcomes they can and cannot provide.} For instance, it is difficult to adjudge in the early stages of ML classification if it's a feasible solution for a given problem, as the ML models tend to be highly dependent on the type and quality of the training data
    ~\cite{sculley2015hidden}.
\end{enumerate}
}

\section{Mapping Scientific Literature and Industry Guidelines}~\label{sec:Checklist}

We developed a mapping of the existing human-centered industry guidelines for building AI software with the findings of a recent RE4AI SLR~\cite{ahmad2021SLR}. We then used this mapping as a baseline for our practitioner survey design, discussed in Section~\ref{sec:Survey}.  

To construct this mapping, we examined four sets of industrial guidelines for building AI software, with three focusing on human-centered AI. Specifically, we analyzed Google PAIR's guidebook on designing people-centered AI systems~\cite{GooglePair2019}, Apple's human interface guidelines for developing ML applications~\cite{Apple2020}, and Microsoft's guidelines for human-centered AI interaction~\cite{Microsoft2022}. We selected these guidelines due to their dedicated focus on human-centered AI software design and development. We chose to use the guidelines presented by Google, Apple, and Microsoft, as they had already done a comprehensive search to provide these guidelines. For example, the paper from Microsoft research involved over 150 AI design recommendations collected from research and industrial sources.  

The fourth set, the Machine Learning Canvas~\cite{MLCanvas}, is a tool that provides startup projects to plan and manage their ML software. The ML Canvas outlines a template for designing and documenting ML systems. Although ML Canvas is not directly targeted at human-centered AI, we decided to include it due to its relevance to our study and the fact that ML Canvas complements the other three. We note that none of the industry guidelines are directly aimed at RE alone but instead at the overall development process. Nevertheless, they provide practical information for requirements engineers and other relevant stakeholders to build human-centered AI software. 

On a closer review of each paper selected in the SLR, we determined that only limited studies focused on human-centered approaches. Of the total 43 studies covered in the SLR, we selected 12 relevant studies for further analysis. We had two conditions for selecting the papers from the SLR: The first one was that the study included human-centered aspects. The second condition was if the paper was relevant to the six selected areas. For example, we included a study that presented requirements for data quality and provided further support to data needs \cite{challa2020faulty}. We also included papers that were secondary studies found during the SLR search. Paper selected from the SLR included: \cite{sandkuhl2019putting, dimatteo2020requirements, schoonderwoerd2021human, krause2016interacting, vogelsang2019requirements, kondermann2013ground, shin2019data, challa2020faulty, ries2021mde, bonfe2012towards, kohl2019explainability, cysneiros2018software}. These papers were mapped against the industrial guidelines discussed in Section~\ref{sec:Background} into six different areas as shown in Table~\ref{table:Matrix}.

\begin{table}[h!]
\caption{Industry/academic map showing human-centred approaches when building AI software}
\label{table:Matrix}
\begin{tabular}{ | p{0.1cm} | p{3.1cm} p{2.2cm} p{3.1cm} p{2.3cm} p{3.2cm}|}
  
\hline  & \textbf{Google PAIR} & \textbf{Microsoft Guide} &
\textbf{Apple Guidelines} &
\textbf{ML Canvas} &
\textbf{Research Papers} \\ \hline 

\multirow{1}{*}{\rotatebox{90}{\textbf\small{User Needs}}}& \footnotesize{ 
• Identify the users need, and if AI is beneficial.
• Decide on automation vs augmentation
• Design, evaluate and monitor reward function (Precision or Recall)}
 & \footnotesize{ 
• Make clear what the system can do, and how well can it do~it
• Provide context based service  
• Remember recent interactions }
 & \footnotesize{ 
• Critical vs complementary (fully or partially) 
• Proactive or Reactive? (Does the user request interaction?)
• Visible or Invisible features (are the users aware of the AI feature)}
 & \footnotesize{ 
• What is the system used for?
• Why is it important?
• Who is going to use it?}
 & 
\footnotesize{ 
• Understand limitations and capabilities of AI \cite{sandkuhl2019putting}
• Trade-off between precision and recall \cite{dimatteo2020requirements}
• Who are the users? What tasks will they perform? what benefits will they gain from using it? \cite{schoonderwoerd2021human}}
 \\  \hline

\multirow{0}{*}{\rotatebox{90}{\textbf\small{Model Needs}}}&\footnotesize{ 
• Model should address users needs
• Tune model with feedback from users
• Balance between overfitting and underfitting
• Improve model to align with implicit and explicit feedback}
&\footnotesize{ 
• Learn from user behavior. 
• Update and adapt cautiously. }
& \footnotesize{ 
• Dynamic: improves model constantly with users interaction and feedback  
• Static: only improves with system updates?}
& \footnotesize{ 
• ML task to use
• When to update?
• How are predictions used to make decisions? 
• Predictions based on new input
• Methods to evaluate predictions?}
& \footnotesize{ 
• Specify what the algorithm should optimize for? \cite{krause2016interacting} 
• Model type (Supervised - requires high quality data, unsupervised - finds patterns in data, reinforced – relies on reward signals) \cite{vogelsang2019requirements}}
\\ \hline

\multirow{0}{*}{\rotatebox{90}{\textbf\small{Data Needs}}}& \footnotesize{ 
• Deciding on features and labels 
• Finding a responsible data source
• Comply with privacy and safety laws
• Avoid and mitigate biases when collecting data
• Design for incoming data from raters and feedback}
& \footnotesize{ 
• ``Match relevant social norms" 
• Mitigate social biases}
& \footnotesize{ 
• Type of data source (public vs private)
• Protect private information and user privacy}
& \footnotesize{ 
• Raw data to use?
• Where is data coming from (internal or external)
• What features are going to use?
• Labelling data:  Explicit vs. implicit }
&\footnotesize{ 
• Data types, amount? Constraints: cost, time, accuracy, quality~\cite{kondermann2013ground}
• Sample rate \cite{shin2019data}
• Data quality:~accuracy, consistency,  credibility, currentness, completeness \cite{challa2020faulty}
• Datasets structure: types, attributes, properties \cite{ries2021mde}}
\\ \hline
\multirow{0.5}{*}{\rotatebox{90}{\textbf\small{Feedback \& Control}}}& 
\footnotesize{ 
• Understand when and why users give feedback
• Explain feedback and impact on system improvement
• Allow users to take control when needed
• Allow users to adjust preferences}
& \footnotesize{ 
•  Plan for consistent feedback 
• ``Support efficient invocation" 
•  Allow feedback dismissal
• Allow user to take control}
& \footnotesize{ 
• Implicit: user interaction
• Explicit:  only ask when required (negative feedback)
• Explain feedback when needed
• Provide multiple options
• Use most recent feedback
• Calibration.}
& \footnotesize{ 
• Methods and metrics to evaluate the system after deployment, and to quantify value creation.}
&\footnotesize{ 
• Allow the user to take over when action is required \cite{bonfe2012towards}}
\\ \hline
\multirow{0}{*}{\rotatebox{90}{\textbf\small{Explainablility \& Trust}} }&\footnotesize{ 
• Explain to not over trust AI
• Explain in general, not specific functionalities.
• Explain how predictions are based on data.
• Explain outputs and predictions
• When to display confidence? 
• Account for user expectations for human-like interaction}
&\footnotesize{ 
• Explain why the system did what it did 
• Convey the consequences of user actions 
• Notify users about changes }
& \footnotesize{ 
• Explain how information is shared between apps 
• Explain benefit rather how it works
• Display Confidence in terms users understand
• Avoid showing low confidence
• Explain what they system cannot do, and inform users when limitation are resolved}
 & 
 &\footnotesize{ 
 • Explain capabilities and limitations of AI \cite{sandkuhl2019putting}.
• Explainability can conflict with requirements as security, cost and precision. \cite{kohl2019explainability}.
• What type of explanations to provide? Adapt explanations to specific context? \cite{schoonderwoerd2021human}
• Transparency to promote trust \cite{cysneiros2018software}}
 \\ \hline

\multirow{0}{*}{\rotatebox{90}{\textbf\small{ Error\& Failure}}}&
\footnotesize{ 
• Define errors and their source
• Account for failstates 
• Identify high stake errors
• Get feedback from rejected predictions 
• Lookout for abusive users}
&\footnotesize{ 
• ``Support efficient correction" 
• ``Scope services when in doubt" (Provide more suggestions)}
&\footnotesize{ 
• Allow users to fix mistakes
• Mitigate effects of mistakes
• Be careful of incorrect assumptions made based on sensitive or private data (leads to mistrust)}
&
&

\\ \hline
\end{tabular}
\end{table}

\subsection{Mapping of the guidelines into the Areas of Human-centered AI}~\label{subsec:mapping}
\textcolor{black}{The six areas were derived from the classification provided in Google PAIR as a baseline. We altered the naming from the Google PAIR chapters (in some cases) to better align with RE terminology, as follows:
\begin{itemize}
    \item Chapter one from PAIR (``User Needs \& Defining Success'') is termed as \textit{Area \#1 -- User Needs} in our mapping.
    \item Chapter three (Mental Models) has been replaced by \textit{Area \#2 -- Model Needs}. Chapter three focuses more on design and the user experience. We added a new area of model needs, which is inspired by chapter three but focuses primarily on capturing the human-centered approaches used to select and train an appropriate AI model. Furthermore, we did not include any sections from any industrial guidelines that focused on the user interface design, as our study is scoped to RE. 
    \item Chapter two (Data Collection \& Evaluation) is termed as \textit{Area \#3 -- Data Needs}. Chapter two from Google PAIR and other industry guidelines focused on how the data was collected and used. In contrast, the mapped data requirements from the SLR focused more on the quality and structure of the data. Thus we replaced the name with ``data needs'' to be more inclusive and encompass both the data collection procedures and other aspects of the training data.
    \item Chapter five (Feedback \& Control) is termed as \textit{Area \#4 -- Feedback and User Control}. We focus on human-centered aspects of the AI software and thus wanted to explicitly focus on the user's control of the system.
    \item Chapter four (Explainability \& Trust) is as-is \textit{Area \#5 -- Explainability and Trust}.
    \item Chapter six (Errors \& Graceful Failure) is termed as \textit{Area \#6 -- Errors and Failure} in our mapping.
\end{itemize}
}
The resulting areas of human-centered approaches and the mapping to different industry and RE literature guidelines are shown in Table~\ref{table:Matrix}. We explain each category below.

\textit{Area \#1 -- User Needs:} 
The first area focuses on capturing the key user needs when building AI systems. This includes identifying the users, their needs from the system, the system's capabilities, and user's interactions with the system. Once identified, the user needs should be analyzed to make crucial decisions, such as whether it is worth building an AI solution. Next, how the user interacts with or views, the system should be specified. Proactive systems interact with the user without requesting it, and reactive systems provide results when people request them or due to the users' interaction with the system and determining if the user is aware of the AI component. Another critical aspect of user needs is determining if the system will augment or automate the users' tasks. Similarly, choosing \textcolor{black}{the evaluation metric (or the `reward function' according to the Google PAIR) should be selected appropriately depending on the user's needs. Based on the type of problem at hand and the kind of AI algorithm being used, the stakeholders should be aware of the choices and the implications of these choices for the users. For example, what would be the trade-off between accepting a false positive vs a false negative in a classification problem? Or, what level of cohesion (or separation) of clusters is acceptable for the users in terms of the task at hand for an unsupervised learning problem.}  

\textit{Area \#2 -- Model Needs:} 
Model needs initially focused on providing an algorithm that would optimize stakeholders' needs. Do you need a system that is explainable or accurate? For example, the study in \cite{krause2016interacting} portrays that some algorithms are better at providing reliable predictions but are difficult to explain. In comparison, other algorithms give better explanations for predictions but lower confidence. Model needs related to tuning and training were mainly connected to data and feedback. Documenting incoming data used in tuning the model should be specified to help mitigate biases. When would this incoming data affect the model? Would it improve while the user interacts with it or only with an update? Will we train the system offline (static) or allow it to update online and learn from the user behavior (dynamic)?  

\textit{Area \#3 -- Data Needs:} 
We identified the key factors that should be considered to establish data needs for human-centered AI systems. These include identifying data sources, types, quality, sampling rate, labels, features, accuracy, and correctness. Setting which data would be used in training and testing the model and identifying how to use feedback in tuning the model should be identified early when building AI software. More importantly, ensuring that the data used is fair and sufficiently inclusive. Data selection should include identifying what features, labels, and sampling rates are needed. Having more samples in the dataset provides diversity when it comes to the data selection~\cite{shin2019data}. However, including more data sets would also increase costs. In this case, one needs to determine the trade-off between the data required to model the AI system to meet user needs and beyond (given the project time and budget constraints) vs. the cost of acquiring more data.

\textit{Area \#4 -- Feedback and User Control:}
Feedback can include asking the user or stakeholders directly for explicit feedback through surveys and ratings or obtaining information from the user's interactions and behaviors with the system through implicit feedback, for example, how many times a day the user accessed the system. Another form of feedback is calibration, which occurs during the initial settings of the system, such as scanning the face to activate face ID.    Plan for different ways to provide feedback, and allow the user to choose which method they want to use to provide feedback. There can be levels to how much of the system the user can take over. In some cases, control can be as simple as providing more than one option to choose from or offering the user choices to adjust their preference. 

\textit{Area \#5 -- Explainability and Trust:}
Most guidelines provide realistic expectations of the AI model to users and stakeholders to help them avoid over-trusting the system and understand its outcomes or decisions.   Explaining data should include how it's shared between other applications, who can access it, and how personal information is stored and used. Predictions can be explained with either confidence levels, an example, or not providing an explanation at all, as sometimes showing confidence could lead to miss-trust. In situations where confidence is low or risks are involved, providing predictions should be avoided. Explaining how feedback is used to improve the model and its impact on the AI system is one of the key components to ensuring user trust. The last of calibrating user's trust involves explaining special cases that might include following laws, rules and orders, and the involvement of third parties. 

\textit{Area \#6 -- Errors and Failure:}
The different error types users might encounter include background errors and user-perceived errors such as context and failstates. Background errors are invisible to the user. Context errors are incorrect assumptions that the system makes about a user and are most likely to be true positives. Failstates, are when the system cannot provide a prediction that it should (true negative) or because of a system limitation. The next step involves identifying error sources and planning how to mitigate them. These include incorrect predictions, data, input, output, and system errors.  

\subsection{Analysis of Key Gaps in Current Industrial Guidelines and Literature}~\label{sec:GapsFromMapping}

After mapping the human-centered AI guidelines and literature, we observed gaps between the industry and research literature. We found that the guidelines were similar for Google, Microsoft, and Apple but relatively different in the literature.  \textcolor{black}{For instance, in data needs (Area\#3), the industry guidelines focus on the type and diversity of data, including the bias in data. The research literature focused more on the quality and quantity of the selected data but not so much on the data bias. In Area\#1, the literature and ML Canvas only mentioned the first part of the guidelines, which was identifying if AI was a valid solution and determining the users and purpose of the system. However, the guidelines went into more detail to identify how the user should interact with the system, user awareness of the AI component, the diversity and inclusion of different types of users, and how much of the system the AI component should control. We did not find any relevant work in RE4AI literature for identifying errors (Area\#6) and feedback~(Area\#4). Overall, in addition to the nine outstanding challenges discussed in Section~\ref{sec:IssuesInRE4AI}, we identified two limitations in our mapping that were not covered in the RE4AI literature, namely, data bias and inclusive design.}

The existing ML canvas framework did not include either explainability or error guidance. We found that the ML canvas focused more on the prototyping phase and did not include any aspects of explainability or user trust (Area\#5). However, it provided an excellent platform to connect requirements engineers, ML specialists, and data scientists.   Finding model needs (Area\#2) from a human-centered approach was challenging, as none of the guidelines specified what methods were required when choosing a model. We could identify a few aspects from Google and Apple's guidelines and split them into this area. We also found a few of the results in the SLR that examined using algorithms in ML systems and how to optimize them to fit user's needs. The ML Canvas focused mostly on the model, and we tried to link them to the human-centered approaches suggested. In the next section, we conduct a survey to try and find which aspects of these guidelines should be addressed in RE. For example, should identifying errors become part RE4AI?

 \section{Practitioner Survey}~\label{sec:Survey}
\vspace{-0.3em}
We describe our survey design and its results on human-centered RE4AI practices. The survey questions were designed using the SLR~\cite{ahmad2021SLR}, literature analysis,  and our mapping of the industrial guidelines to literature. We wanted to identify what current modeling languages, tools, and techniques are used. And the limitations and challenges faced during the RE phase when building AI software. We also aimed to identify human-centered approaches used in current RE practices when designing and building AI software. 

\subsection{Survey Design \& Recruitment}~\label{subsec:surveydesign}

When designing the survey, we followed Kitchenham and  Pfleeger's~\cite{kitchenham2002design} guidelines to principles of survey research. The \textcolor{black}{survey questions}\footnote{\href{https://www.dropbox.com/s/a5h1rzknel9zod6/Survey_Questions.pdf?dl=0}{Survey Questions Link}} covered three sections: 
\begin{enumerate}
\item The first section covered demographics regarding the participant's background knowledge, including their role in the organization, years of experience while working on projects that involve building AI software, experience with AI components, and the application domain of experience.
\item  The second section aimed to identify requirements and modeling tools used in RE4AI practices. We also wanted to know what challenges and limitations people face when specifying requirements for AI systems. We used the nine issues presented in the \textcolor{black}{literature (Section~\ref{sec:IssuesInRE4AI}) and included two more that we identified from the guidelines (Section~\ref{sec:GapsFromMapping}). The additional issues were regarding identifying biases in data and issues with providing inclusive designs~\cite{amershi2019guidelines}. We asked a question in our survey \textit{``What are the challenges or limitations you face when specifying requirements for Artificial Intelligence (AI) systems?''}. We displayed these eleven outstanding challenges in RE4AI literature in a list format as potential responses to this question, and each participant could select more than one choice. We wanted to understand if the survey participants had encountered these issues in practice. We also gave them a choice to enter other issues they had encountered.}
\item The third section targets which human-centered approaches are needed when building AI systems in RE. These questions were designed based on the human-centered guidelines gathered in Section~\ref{sec:Checklist}.  \textcolor{black}{We summarized the guidelines from the mapping of Table~\ref{table:Matrix} and created a bullet point list for each area in a question format (see Figure~\ref{fig:human-centred-Needs}). The participant could select the items they thought should be or are addressed in RE. The given codes for each area are used with the results in Figure~\ref{fig:HumanNeeds}. For example, U1 is the code for ``identify the user needs'', and in Figure~\ref{fig:HumanNeeds} we only use U1 instead of the entire text. 
For Areas\#1-\#6, we wanted to determine which human-centered approaches were used in RE. Also, we wanted to know how many of our participants did not consider the human-centered approaches when building AI software solutions. To get this information, we provided an option for each area from Areas\#1-\#6, e.g., ``Our projects do not specify any data requirements in the Requirements Engineering stage''.
Therefore, we left a selection for each area where the participant could select that their projects did not specify any human-centered needs in RE. 
}
\end{enumerate}

We created the survey using the Qualtrics platform and designed it to be compatible with computers and mobile phone platforms. At the start of the survey, a plain language statement was provided to explain the research impact gained by the participant's involvement. Before publishing the questionnaire, we conducted a pilot test with three experts in the field of RE4AI. The pilot study helped us find ambiguous questions, the time it took to complete the survey, and decide on questions that might identify biases. The survey was then published using the Qualtrics platform, and we provided the link and a QR code to participants.  \textcolor{black}{We used a combination of convenience sampling and snowballing methods to recruit participants. The convenience sampling method was used by forwarding the survey to our professional contacts who met the inclusion criteria for the survey (see Table~\ref{table:inclusionExclusionSurvey}). We then used snowballing by asking those contacts to forward the survey to anyone who fit the criteria. To reduce the selection bias, we further advertised the survey on platforms such as LinkedIn, Twitter, and Reddit.} The criteria for the target population were software engineers, requirements engineers, data scientists, machine learning specialists, and project managers that have been building or managing software systems with an AI/ML component. The survey was completely anonymous, no incentive was provided, and participants took approximately 5 to 15 minutes to complete it.   

\subsection{Selection Criteria}

For selecting the final responses, we set the selection criteria as shown in Table~\ref{table:inclusionExclusionSurvey}. 

In total, we had 65 people consent to the survey. 
Only 29 complete responses were submitted. 12 participants only selected the consent question, and 24 of the responses were partial. We did not use partial responses as we specified in the participant information form that only submitted results would be used in case the participants wanted to withdraw from the study. 
To gauge participants' experience in building AI components or systems, we had two questions asking about their experience with AI. The first one was ``Which of the following AI capabilities have you worked with?''. The second question asked how many years of experience they had working with AI. We checked all 29 responses to validate their experience working with a type of AI software solution. 

\begin{table}[htbp]
\textcolor{black}{\caption{Initial inclusion and exclusion criteria for the selected survey results}}
\label{table:inclusionExclusionSurvey}
\begin{tabular}{p{6cm} p{6cm}}
\hline
\textbf{Inclusion criteria:}  & \textbf{Exclusion criteria:} \\ \hline
The consent question is selected  & Consent question is not selected \\ 
Completed surveys & Partial responses \\
The participant has experience working with AI components or systems. & The participant has no experience working with AI components or systems.\\ \hline
\end{tabular}
\end{table}  

\subsection{Results}~\label{subsec:results}

In total, we had 29 selected responses. The final results had 20 males, six females and three preferred not to identify their gender. We asked participants how many years of experience they had working specifically on AI projects. Nine participants had less than three years of experience working with AI, 11 participants had between three to five years, and the rest had over six years of experience. These results show that most of our respondents had less than five years of experience in using AI, which is reflective of the recent developments in the AI field. 

\begin{figure}[h!]
   \centering
\includegraphics[width=0.6\linewidth]{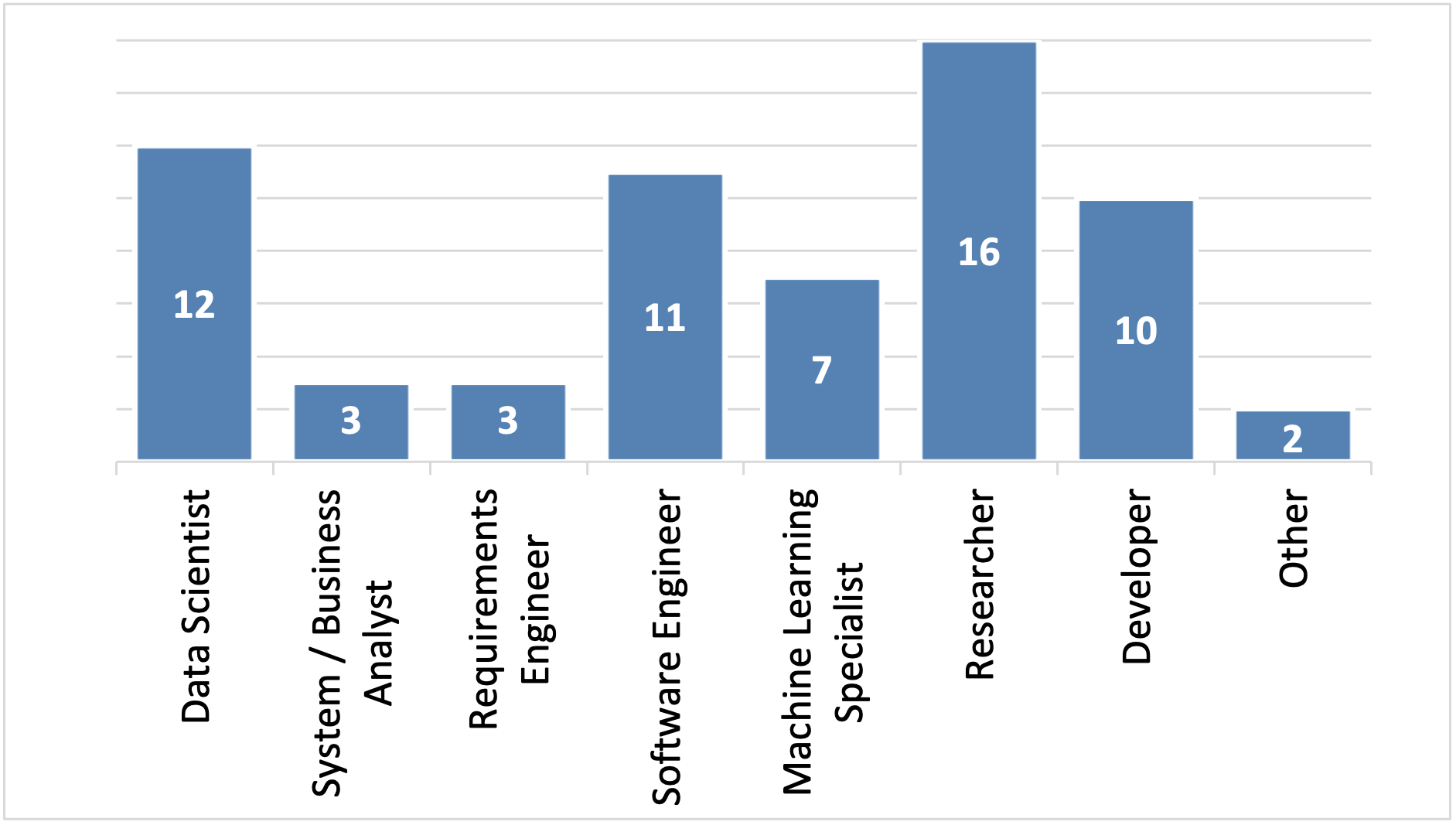}
\caption{Different roles of participants in the organization}
\label{fig:Role}
\end{figure}

\textcolor{black}{We asked about the (industrial) role(s) each person had. Participants were able to select more than one choice. Some reported having two or more roles. The rationale behind the ability to select multiple roles was to capture the experience of participants in AI projects from their current or past roles.
For example, some researchers reported that they were also data scientists. This could mean three things - (i) either they have worked in the past as a data scientist on an AI related project; (ii) means either they have worked as a researcher on AI projects in the past, and currently work as a data scientist (or the other role reported); and (iii) they are working in the industrial research and development and have concurrent roles.
16 people reported to be researchers, 12 of them reported with at least one more role other than the researcher, and four reported as only researcher. 
Data scientists were the next popular participation group, with 12 participants, followed by 11 software engineers and 10 software developers. Finally, we had only three people reporting to have worked on RE-related roles, and three other people reported their roles as business analysts.
}

\subsubsection{Application Domain and AI Tasks}

We asked participants about the application domain in which they have applied AI or developed AI solutions. Participants could select multiple domains as some might have changed work environments or made solutions for multiple organizations. Our results had education as the highest domain with 14 - this might be linked to the fact that our highest responses were from researchers. The next most popular domain was building AI solutions for governmental organizations at six responses. Followed by the defense at four responses. Manufacturing, automotive, health, and banking with three responses each. Two participants said they built AI solutions for the entertainment industries. And finally, one response for each agriculture and food, and retail. 

\begin{figure}[h]
   \centering
\includegraphics[width=1.05\linewidth]{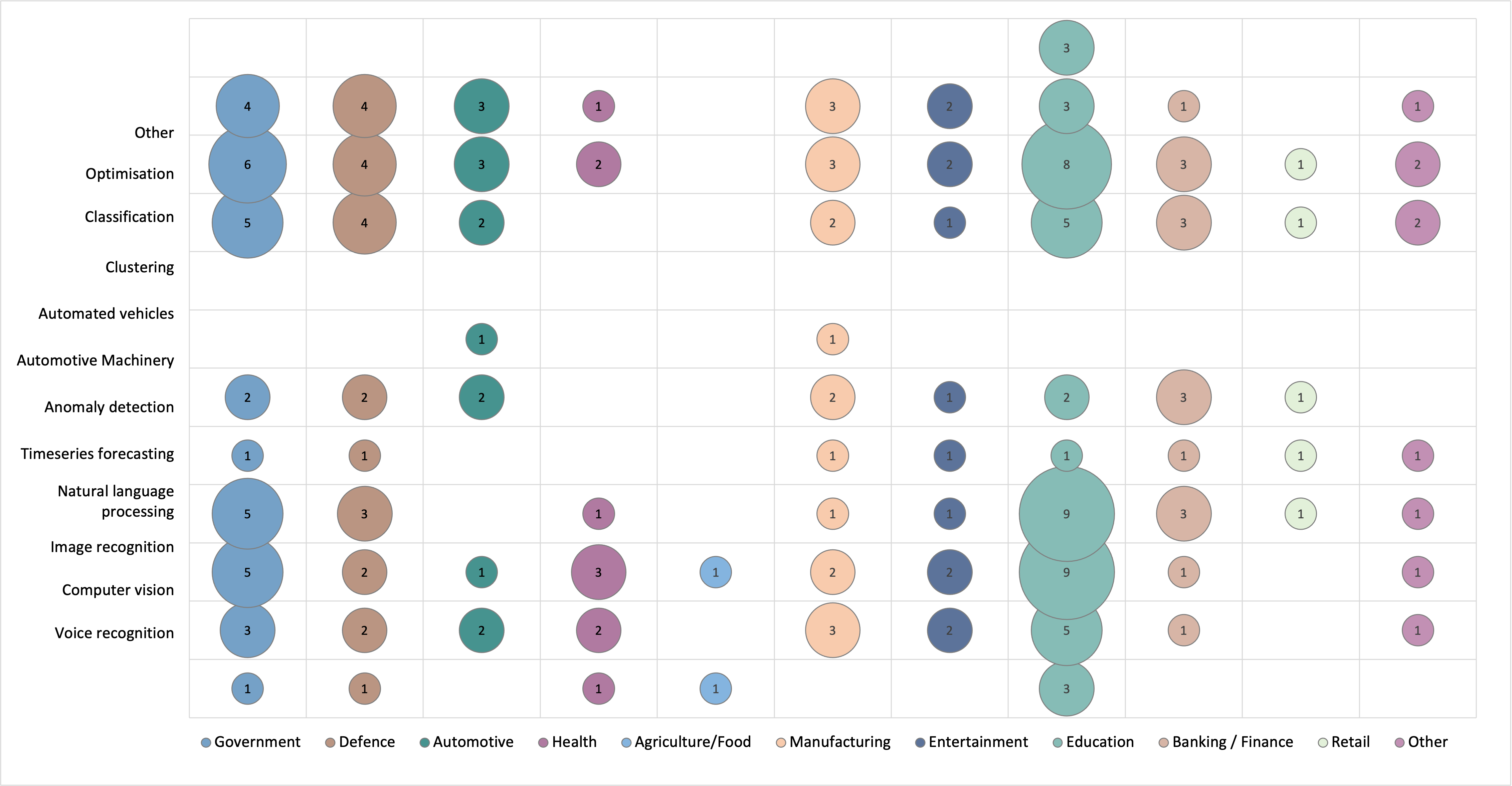}
\caption{Mapping of the application domains in the industry with the different AI-related tasks.}
\label{fig:ApplicationDomains}
\end{figure}

We further asked participants about the AI tasks they have worked on or used more often in their application domain. AI tasks, such as Natural Language Processing (NLP), image recognition, and classification, received the highest responses among our participants. Computer vision, clustering, and optimization came next. Figure~\ref{fig:ApplicationDomains} shows the mapping of each of these AI tasks with each application domain. We note that none of the 29 participants had worked on automated vehicles. The results are contrasting to the RE4AI SLR, wherein the automated vehicles received the highest interest~\cite{ahmad2021SLR}.  

\subsubsection{Types of Requirements Used}~\label{subsec:reqTypes} 

Table~\ref{table:ReqType} shows the participants' responses to the requirements types they work with during their AI software development projects. Only three participants did not include RE in building their AI software.  
The majority of participants reported to focus on functional requirements for both AI and non-AI components of the system.
Non-Functional Requirements (NFRs) were included in building AI software but did not receive as much attention as functional requirements. Data requirements were the next most frequently specified. 15 participants reported on specifying some sort of requirements to deal with data collection and management. Ethical requirements seemed to have the lowest response rate among the type of requirements specified. \textcolor{black}{The low coverage of ethics requirements could be due to several plausible reasons, such as ethics are not considered as important in practice, such requirements are not considered due to time and budget pressures, ethics are considered to be high-level system objectives that do not need to be specified with system requirements~\cite{lu2022software}, or people do not have an understanding of the ethics requirements. Further investigation is required on this to determine the role of requirements related to AI in practice.}


\begin{table}[h!]
\centering
\caption{The different types of requirements used by participants} 
\label{table:ReqType}
\begin{tabular}{p{7cm} p{0.7cm} }
\hline
\textbf{Requirements Type} & \textbf{Count} \\ \hline
Traditional (non-AI) functional requirements & 16    \\ \hline
Traditional non functional requirements & 8     \\ \hline
AI System functional requirements & 21    \\ \hline
AI System non functional requirements & 7     \\ \hline
AI System data requirements & 15    \\ \hline
AI System user experience \& interface requirements   & 7     \\ \hline
Ethical requirements & 6     \\ \hline
No requirements specified when building AI & 3     \\ \hline
\end{tabular}
\end{table}

\subsubsection{Tools Used} 

\begin{figure}[h]
   \centering
\includegraphics[width=0.85\linewidth]{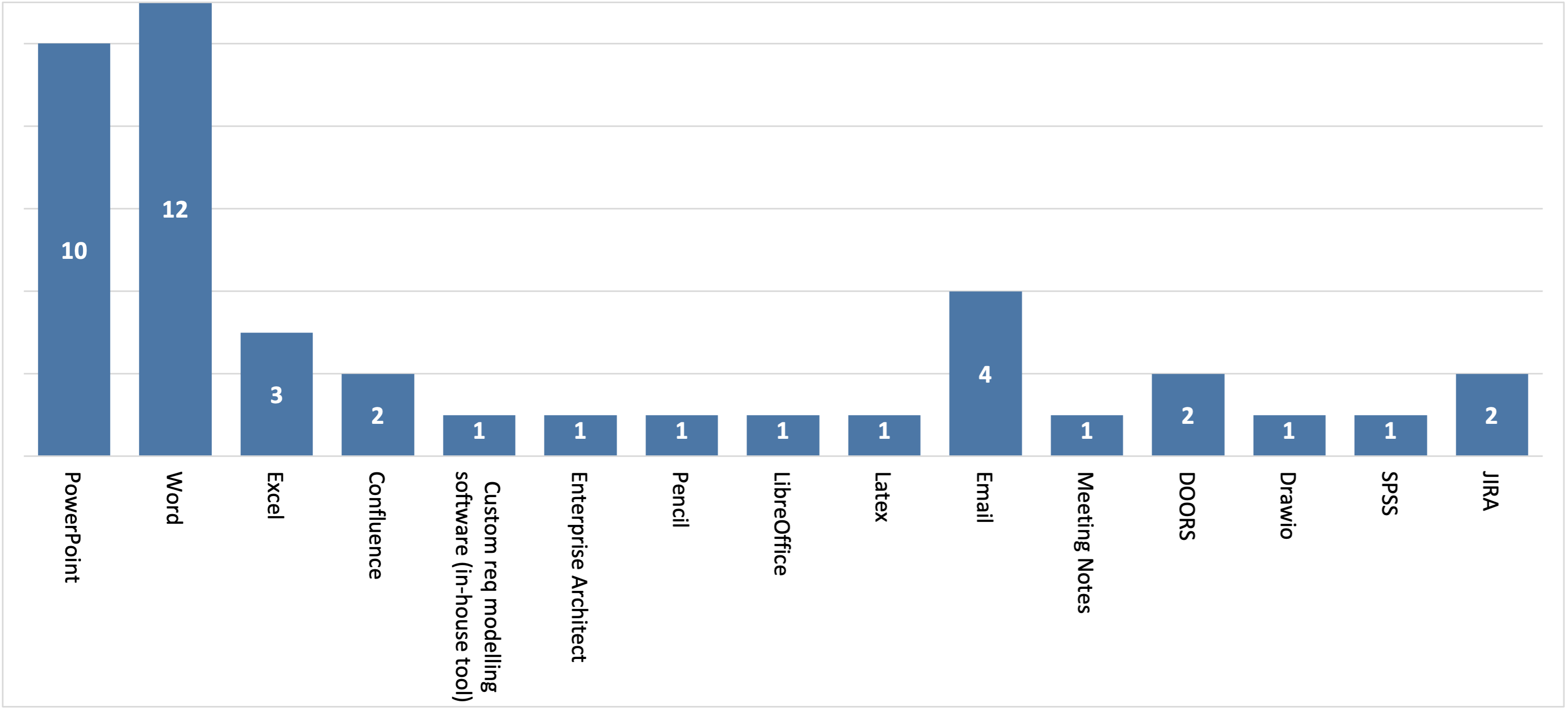}
\caption{Tools used by participants and the number of times they were mentioned}
\label{fig:Tools}
\end{figure}

We identified 16 different tools that were used in writing and managing requirements for AI systems, as shown in Figure~\ref{fig:Tools}. Commercial software for planning and documenting software solutions such as JIRA~\cite{Jira} and Confluence~\cite{Confluence} were used by two participants each. JIRA and Confluence are software applications used for agile project management. One participant used SPSS~\cite{SPSS},  a commercial software solution developed by IBM solutions for statistical analysis. DOORS~\cite{DOORS} was used by two participants in managing requirements for AI systems. DOORS is a software solution created by IBM specifically built to optimize and manage requirements. 

Collaboration-supporting platforms included the free open-source software LibreOffice~\cite{libreoffice}. Emails were used by four participants to communicate requirements. Drawing platforms and modeling tools such as Enterprise Architect~\cite{EA}, Pencil~\cite{Pencil}, Drawio~\cite{Drawio} were mentioned once each as methods to display requirements visually. Enterprise Architect is a UML modeling tool. Pencil is a free open-source GUI tool to aid in drawing images and shapes. And Draw.io is an online design platform that is used to draw flowcharts, UML diagrams, ER diagrams, etc. 

\begin{figure}[h]
   \centering
\includegraphics[width=0.5\linewidth]{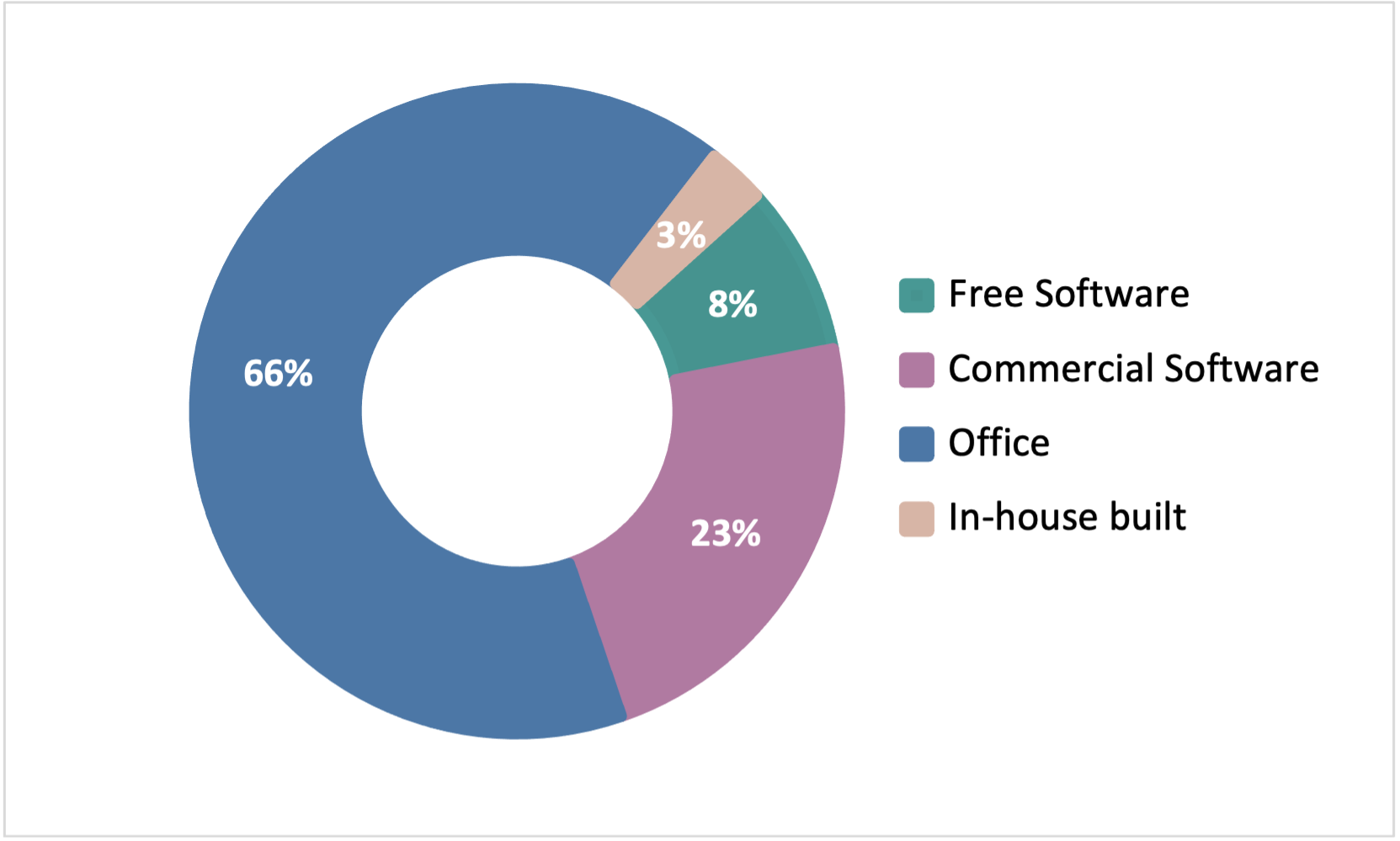}
\caption{The different software platforms for tools used by participants}
\label{fig:ToolsPlatform}
\end{figure}

Platforms such as Microsoft Office were the most popular method used to document requirements. Word was used by 12 participants, followed by 10 using PowerPoint and three using Excel. Microsoft Office making up 66\% of the total tools used. We noticed that commercial software was more popular than free and open-source software. And one of the participants used a custom in-house tool they developed to model requirements, as shown in Figure~\ref{fig:ToolsPlatform}.

\subsubsection{Modeling Notations and Methods Used to present requirements}

\begin{figure}[h]
   \centering
\includegraphics[width=0.6\linewidth]{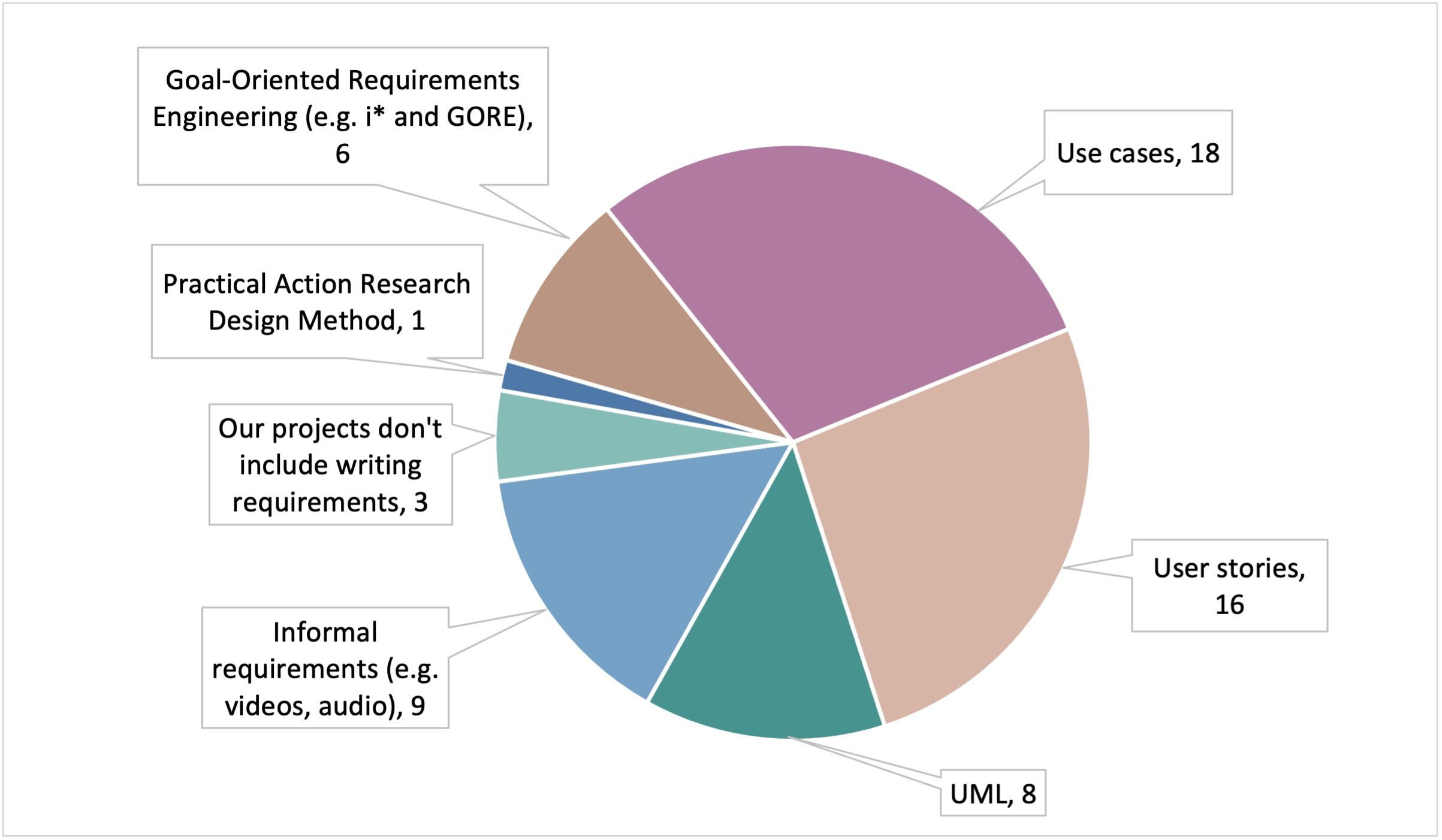}
\caption{The different modeling languages and notations used by participants to specify requirements for AI.}
\label{fig:ModelingNotations}
\end{figure}

The survey revealed that use cases and user stories were the most popular method among our participants to elicit requirements in our survey, with 18 participants using use cases and 16 for user stories, as shown in Figure~\ref{fig:ModelingNotations}. User stories came next with 16 selections, followed by informal requirements such as video and audio recordings at nine. Modeling notations were not as popular to display requirements. Only six participants used Goal-Oriented Requirements Engineering (GORE). Unlike UML, GORE had better support when it came to modeling NFR's and business rules. However, it is more difficult to learn and use~\cite{neace2018goal,dimitrakopoulos2019alpha}. Given that we only had three requirements engineers in our survey, we found fewer responses from people that said they used GORE. Three of our participants did not consider requirements and went straight into implementing AI software projects. Finally, one of the participants entered Practical Action Research Design Method into the ``Others'' selection.

\begin{figure}[h]
   \centering
\includegraphics[width=1\linewidth]{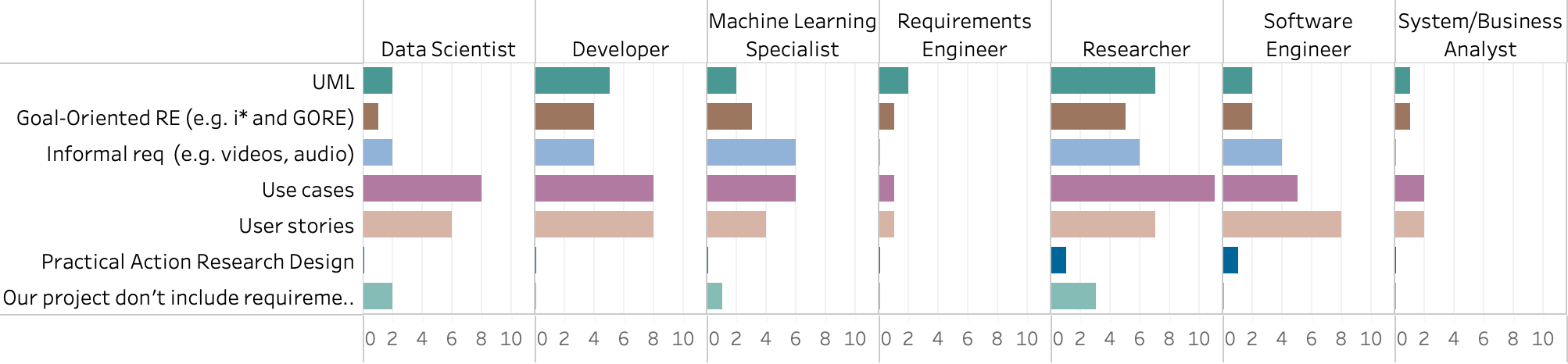}
\caption{Method used by different participants to document requirements for AI.}
\label{fig:ModelingPerRole}
\end{figure}

Use cases were the most popular methods used among the participants. Figure~\ref{fig:ModelingPerRole} shows that the most used methods were use cases and user stories, regardless of what role they had in the building process of AI software. Only one person who was a software engineer researcher used Partial Action Research Design. Researchers and developers preferred using visual presentations such as UML and GRL. 

\subsubsection{Issues with current practices}

\begin{figure}[h]
   \centering
\includegraphics[width=0.9\linewidth]{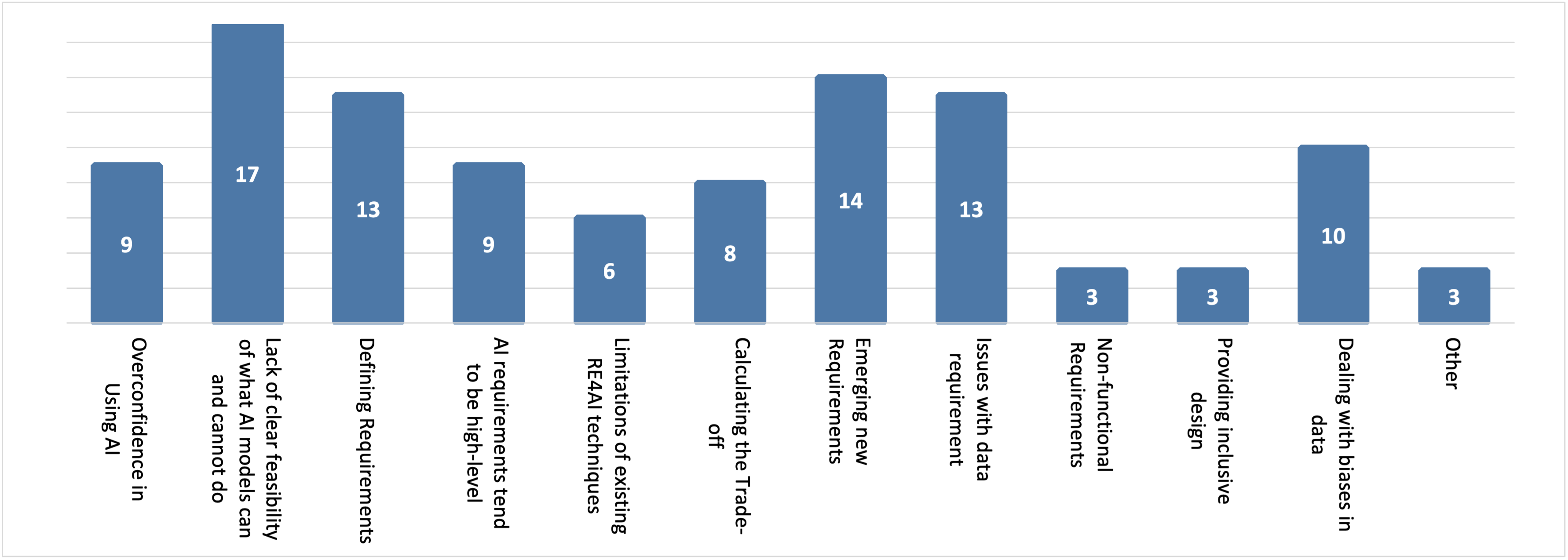}
\caption{Issues recurring in the survey results}
\label{fig:Issues}
\end{figure}

Initially, we gathered the issues found in the literature and identified two new ones from mapping the guidelines, as shown in Figure~\ref{fig:Issues}. Three new issues emerged from our participant's responses to include:

\begin{enumerate}
\item Requirements needed for data collection.
\item Requirements for designing robust interfaces that take into account the issues with non-deterministic systems.
\item Issues with eliciting requirements from clients - as usually, the client does not necessarily understand the capabilities and limitations of AI systems.
\end{enumerate}

The biggest issue was the lack of clear feasibility of what the AI models can and cannot do, which was selected 17 times. With the process of building AI software being vastly different from software that does not have an AI component, new requirements have emerged. So working with some of these requirements, such as data, explainability, and ethics would be used differently in AI software, causing new issues and the need to build new approaches to manage such requirements. This issue was selected 14 times in our survey responses. Defining requirements and issues with data requirements came next; both were selected 13 times. The overconfidence in using AI was identified as an issue by nine participants. Issues related to calculating the trade-off when building AI systems (e.g., precision vs recall or privacy vs explainability) had eight responses. Finally, issues with non-functional requirements and providing inclusive design received the least interest. \textcolor{black}{Similar to our surmise in Section~\ref{subsec:reqTypes} regarding the low coverage of ethics-related requirements, we believe that there could be several plausible reasons for the low coverage of NFRs, including those already covered in Section~\ref{subsec:reqTypes}, the lack of knowledge and expertise or complexity involved in dealing with NFRs, or the costs associated with testing NFRs for AI~\cite{habibullah2021non}.
}

\subsubsection{Human-centred Needs}

\begin{figure*}[h!]
   \centering
\includegraphics[width=1\linewidth]{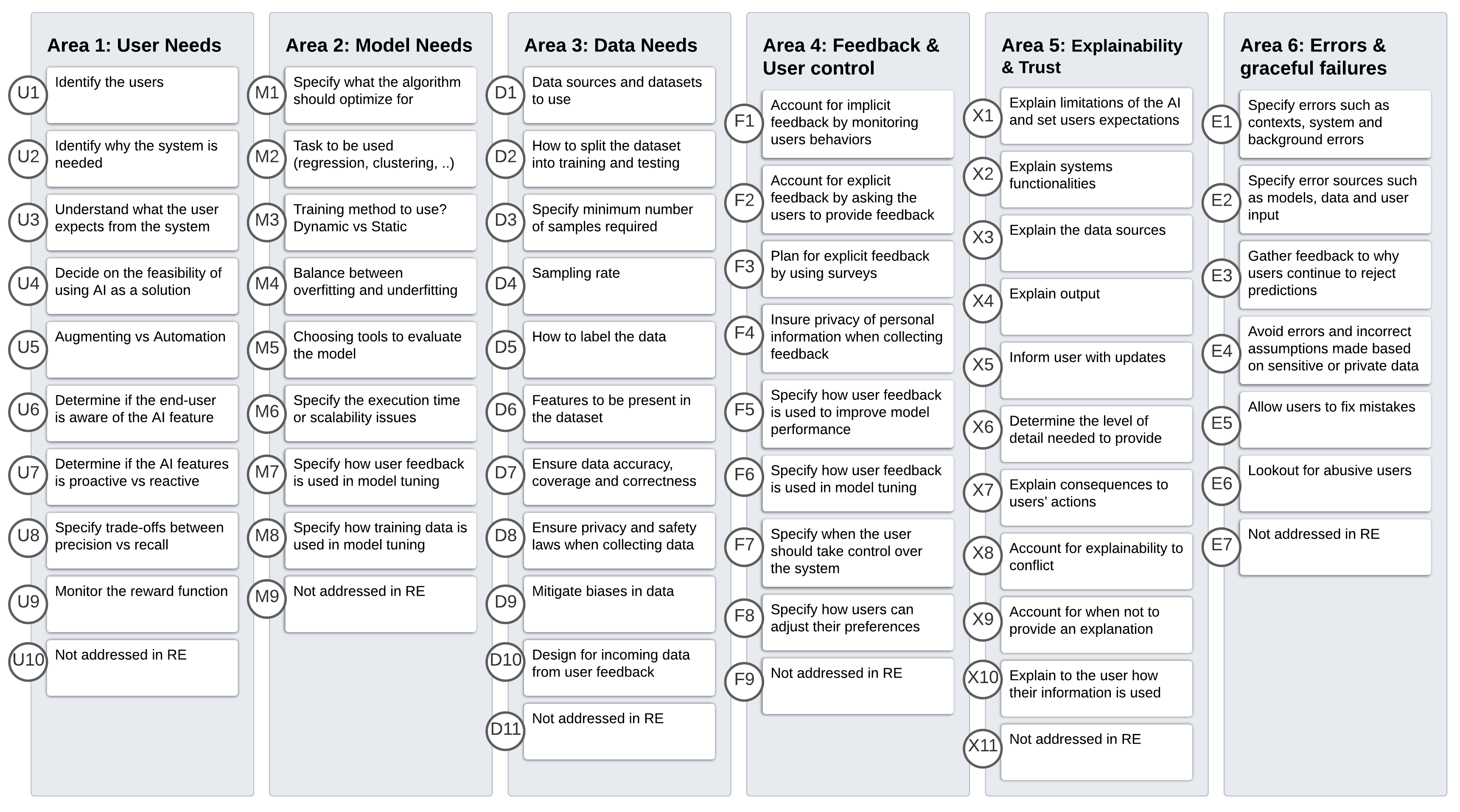}
\caption{Targeted human-centred approaches when building AI systems}
\label{fig:human-centred-Needs}
\end{figure*}

\begin{figure*}[h]
   \centering
\includegraphics[width=1\linewidth]{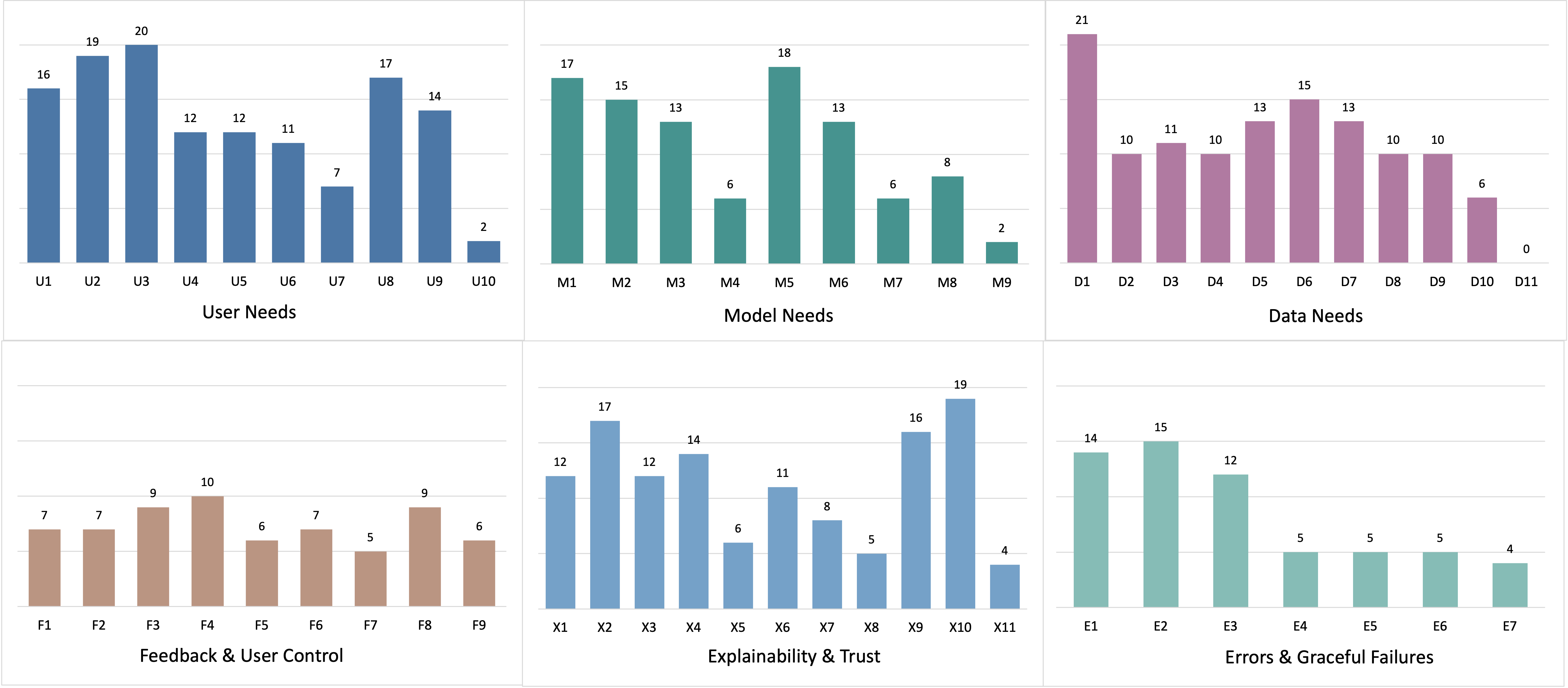}
\caption{Human-centered needs for AI as displayed in Figure~\ref{fig:human-centred-Needs} and the number of times they were selected}
\label{fig:HumanNeeds}
\end{figure*}

In section (iii) of the survey on human-centered approaches for RE4AI, each of the human-centered guidelines from the six areas in Section~\ref{subsec:mapping} was selected by at least one participant. Therefore, we conclude that all human-centered guidelines should be addressed during the RE phase when building AI systems. The code representations for each need are shown in Figure~\ref{fig:human-centred-Needs}, and the number of participants who considered these needs important for RE4AI is shown in Figure~\ref{fig:HumanNeeds}. Overall, user, model and data needs, and explainability and trust received more interest from participants than feedback and user control and errors.

For user and model needs (Areas\#1 and \#2), only two participants in each area responded, as these were not being addressed during the RE phase in their experience. All participants responded that data needs (Area\#3) are considered in the RE4AI projects. Six participants said they did not consider feedback and user control (Area\#4), and four participants each did not consider explainability and trust (Area\#5), and errors (Area\#6) in RE.

From user needs (Area\#1), U2, referencing identifying why the system is needed, and U3, understanding what the user expects from the system, had the highest responses at 19 and 20 selections. Seventeen participants chose U8 to specify the trade-off when calculating the reward function and to determine between precision and recall. Fourteen selected U9 to monitor the reward function over time. In U1 identifying the user needs, got 16 responses.  
Overall, all needs (U1--U9) in Area\#1 (except U7) were selected by more than 10 participants, showing the importance of identifying them during the RE phase. In model needs (Area\#2), M5 -- choosing the tools required to evaluate the model got the most interest. Overall, M1--M3, M5, and M6, received responses from more than 13  out of 29 participants. Followed by M1 specifying what the algorithm should optimize for, whether it should optimize for explainability, security, time,~etc,.
In Area\#3, identifying the data sources and datasets to use when building AI software (D1) was selected by 21 participants making it the one with the most responses. Overall, all needs (except D10) in Area\#3 got a significant ($>10$) number of responses across all areas, emphasizing the importance of specifying data needs while building AI software. Participants also found D6 important, which was ensuring that the data is accurate, covers all the possible groups, and is correct. D8, which was ensuing data privacy and safety, comes next.  
In explainability and trust (Area\#5), 19 people selected X10, which was explaining to the user how their information was used. Other needs (X1--X4, X6, X9) received ($>10$) responses as well.
Feedback and user control (Area\#4) and errors and graceful failures (Area\#6) did not get as many responses as other areas. 
In Area\#4, participants showed more interest in F4, ensuring the privacy of user information when collecting feedback, F3 planning surveys to gather explicit feedback, and F8 specifying how users can adjust their feedback.  
In Area\#6, E1 specifying error types, E2 specifying error sources, and E3 finding out why users reject predictions had more selections than the rest for errors. 
Although Areas\#4 and \#6 had a lower response rate, more than 23 participants still thought they were important to address during~RE.

\newpage
\subsection{Threats to Validity}~\label{sec:Threats}

\textit{Internal validity:} 
We carefully designed our survey based on the guidelines and SLR to reduce internal validity threats that might have affected our final results. We created a protocol to identify how each question will be evaluated and conducted a pilot with three RE4AI experts to address any issues in the survey questions.  
Despite the steps taken above to improve our survey's quality, we acknowledge that we might have missed some of the guidelines that focused on human-centered aspects as we only included the industrial guidelines. Having said that, we found that guidelines such as the ones from Microsoft and Google PAIR already included an extensive search in human-centered AI research, thus improving the level of confidence in the guidelines covered. \textcolor{black}{We further note a potential threat to internal validity was over the consistent understanding of certain concepts used in our survey, e.g., traditional vs AI system requirements (Table~\ref{table:ReqType}). We tried to mitigate this threat by clarifying some of the concepts in the participant information sheet and also provided our contact information to allow the participants to get in touch with us in case of any doubts.
}

\textit{External validity:} 
When conducting the survey, we found that the number of male participants was higher than female participants. However, when comparing the general years of experience, we noticed that each gender group had equally balanced years of experience. We also noticed that we had a higher number of participants that were researchers. But we found that the input obtained from researchers was equally valuable as that from non-researchers. 

\textit{Conclusion Validity:} 
To mitigate conclusion validity threats, a protocol was created prior to writing and conducting the survey to identify how the results would be evaluated. Also, the authors had weekly meetings to discuss data extraction and analysis methods to determine how the results would be used. 

\section{Discussion}~\label{sec:Discussion}
This section discusses the gaps we found when conducting the survey and provides future recommendations. We found some differences in the tools, methods, domains, and issues from existing practices in the literature.

\subsection{Gaps between Literature and Practice}

\textcolor{black}{The advantage that the survey had over the SLR was that we could capture more tools used by people in the industry, as most studies in literature would not mention the use of tools such as text editors and collaboration platforms.} In practice, a larger number of tools were used than the ones reported in the literature. The SLR~\cite{ahmad2021SLR} identified only two tools to include  {jUCMNav} a free graphical editor for modeling Goal-oriented Requirement Language (GRL)\cite{amyot2011grl} and \cite{ries2021mde} built a toolset based on the {Sirius framework} which is an open-source graphical editor for Domain-Specific Modeling (DSM)~\cite{viyovic2014sirius}. On the contrary, the survey showed 15 different tools and drawing editors. We also noticed that most of the software used by practitioners was commercial software, whereas studies in literature only used open-source or free software.   We also found that issues presented in the survey were vastly different from what is available in the literature. The overconfidence in using AI was only identified in one study in the SLR. Whereas nine respondents thought it was an issue in the~survey. 

When comparing the modeling languages, we found that the use of UML was similar in the frequency of use in comparison to the SLR results. We noticed that most of the users were using UML to model requirements. The reason for choosing to use UML was that it was more common among non-software engineers and its ease of use~\cite{ahmad2021SLR}. However, UML had its limitations when it came to modeling NFR's~\cite{neace2018goal,dimitrakopoulos2019alpha}. Although, we also found that there was less emphasis on working with NFR's and more focus on functional and data requirements, as shown in Table~\ref{table:ReqType}, which might have contributed to having more participants preferring to use UML in the survey results. On the other hand, we found that GORE was not as used in the survey results in comparison to the SLR. In the SLR, GORE accounted for 24\% of the total modeling languages. Whereas, in the survey, only 7\% of the participants said they used GORE. The reason could also be linked to the limited number of requirements engineers in the survey responses or could be that GORE is not widely adopted in the industry compared to UML.

Another gap that we observed was in the application domains. In the SLR, most studies focused on Autonomous vehicles and the health domain. In contrast, in the survey, these two domains were less of interest, and most of the applications contributed towards education, government, and defense. None of the studies in the literature addressed these three application domains.

\textcolor{black}{When reporting on our participants' roles, as mentioned in Section~\ref{subsec:results}, we only had three people report that they have worked on RE-related tasks and three others reported having worked as a business analyst. We surmise that this could be the case because: (1) there is limited work on RE4AI, or RE tasks are lacking in the field of AI software development; and/or (2) RE is viewed differently in practice, i.e., it is considered as a knowledge area rather than referring to a specific role in an organization. For example, people have different titles, e.g., business analyst, project manager, and software engineer, while they perform RE-related tasks~\cite{herrmann2013requirements,wang2018understanding,daneva2017job}.}

\subsection{Gaps in Human-centered AI}

The survey provides insight into which human-centered AI guidelines should be addressed in RE; this was not evident in the SLR. When identifying user needs, the first step was to identify the users and if AI is a solution. Surprisingly we found that U4 had only 12 responses, which meant that only 40\% of the participants considered the feasibility of using AI as a solution before using AI. Google PAIR emphasized greatly on the importance of finding if AI is a feasible solution. And a survey in \cite{sandkuhl2019putting} showed that organizations often decide to use AI without identifying the need for it or just because they have the data. We find that this needs to be an important part of RE4AI.   In the survey, only 10 people mentioned that they tried to mitigate biases in data selection. We found this to be a low response as the guidelines stressed identifying key data characteristics at early stages to help reduce biases. Other related reasons for data biases included missing and unexpected features,  under or over represented data, not including minority groups in data collection \cite{GooglePair2019}, human labeled data \cite{vogelsang2019requirements}, and using existing data. Using existing data could make it difficult to explain why a given prediction is provided \cite{kohl2019explainability}. We find it important to identify and report biases in data as early as possible to avoid any issues that might evolve into producing biased AI.  

Setting user expectations in the SLR results and guidelines was emphasized more than explaining how the user information was used. However, when it came to explainability in the survey, we found that X1 - explaining the limitations of AI and setting the user's expectations, had only 10 responses, whereas X10 ``explaining how user information is used" had 19 responses. This meant that in practice, most of the participants felt that it was more important to explain how information is used. None of the studies in the SLR mentioned the need to address errors in the RE phase when building AI systems. However, addressing (identifying and dealing with) errors in RE was greatly emphasized in the guidelines. The survey results showed that people working on AI systems wanted to know how to deal with errors and specify error sources during RE.  

\subsection{Future Recommendations}
We propose the following research recommendations:

\subsubsection{Recommendation 1:} Engineering AI systems introduced new specs that did not exist in traditional software to include data, model specs, feedback, explainability, etc. Therefore, this would require either new tools or extending existing tools. Our survey presented a similar pattern. We found that the industry is using basic tools like JIRA, Excel, etc. Although these tools are flexible and easy to use, they do not enforce or consider any RE4AI attributes in AI specifications. Our hypothesis is that this would lead to poor implementation of AI systems. Thus, we recommend the RE4AI community considers inventing new tools to address these gaps. We need to re-evaluate and research the tools and platforms currently used to manage requirements, as evident from our survey that existing ones are not equipped to handle~RE4AI. 

\subsubsection{Recommendation 2:} We recommend that more work be invested in providing a suitable modeling language to model requirements. We found that most participants were using UML or Microsoft Office to present requirements for AI, which poses an issue with the quality of requirements presented for AI. In its original form, UML is not suited for modeling all requirements for AI systems, such as data, ethics, or other~NFRs. Therefore, modifying or extending existing modeling languages is necessary to support RE4AI.

\subsubsection{Recommendation 3:} Results from the SLR showed that there are limited studies in RE that focus on human-centered aspects when building AI systems. We also found that practices in the industry favored including human-centered AI aspects in RE.   We recommend that the mapping resulting from the human-centered guidelines presented in the paper could be used to create an RE framework to specify and elicit requirements for human-centered AI software.

\subsubsection{Recommendation 4:}  When examining the human-centered AI guidelines, we found that all the industrial guidelines emphasize on identifying errors and linking them to user needs. However, we found no mention of including errors in RE4AI literature. We argue that identifying and managing errors should be included in RE4AI. Once the error types and sources are identified, an action plan needs to be considered. The action plan should consist of steps to be addressed to mitigate and fix these errors. These might include ways the user is allowed to fix these mistakes or provide them with other suggestions that they could use in case an error occurs. 

\subsubsection{Recommendation 5:}  Feedback is an integral part of AI systems, as it could be the defining factor in how well the user interacts with the system and, in some cases, could be used in tuning and training the AI-model.  Similar to recommendation four we found that there was no existing work on feedback in RE4AI literature, even though there was great emphasis on managing feedback in industrial human-centered AI guidelines.   We recommend that identifying the different types of feedback needed from either the user, rater, or stakeholder should be included during RE practices, as well as specifying how the feedback will be used to improve the performance of the AI software.

\section{Conclusions}~\label{sec:Conclusion}

In this paper, we mapped current industry human-centered AI guidelines and existing literature on RE4AI.  The results from the mapping were used in a survey to find currently used approaches in RE4AI and to identify which of the human-centered approaches should be determined during RE when building AI software. We were able to establish that all the mapped human-centered approaches should be included in RE4AI.  We also identified the gaps between literature and practice.  And finally, we mapped the requirements for human-centered AI into six different areas: user needs, model needs, data needs, explainability \& trust, feedback \& user control, and errors.  \textcolor{black}{In the future, we would want to expand on our survey results to build a framework for eliciting and specifying requirements for human-centered AI projects. We would further explore the current industrial practices via interviews with the practitioners.}


\printcredits


\bibliographystyle{elsarticle-num}
\bibliography{mybib}

\begin{thebibliography}{10}
\expandafter\ifx\csname url\endcsname\relax
  \def\url#1{\texttt{#1}}\fi
\expandafter\ifx\csname urlprefix\endcsname\relax\def\urlprefix{URL }\fi
\expandafter\ifx\csname href\endcsname\relax
  \def\href#1#2{#2} \def\path#1{#1}\fi

\bibitem{wilson2018collaborative}
H.~J. Wilson, P.~R. Daugherty, Collaborative intelligence: humans and {AI} are
  joining forces, Harvard Business Review 96~(4) (2018) 114--123.

\bibitem{jiang2017artificial}
F.~Jiang, Y.~Jiang, H.~Zhi, Y.~Dong, H.~Li, S.~Ma, Y.~Wang, Q.~Dong, H.~Shen,
  Y.~Wang, Artificial intelligence in healthcare: past, present and future,
  Stroke and vascular neurology 2~(4) (2017) 230--243.

\bibitem{FacebookChatbot}
{Andrew Orlowski}, Facebook scales back \uppercase{AI} flagship after chatbots
  hit 70\% f-ai-lure rate, the Register.
  https://www.theregister.co.uk/2017/02/22/facebook\_ai\_fail. Mar 22, 2017.

\bibitem{price2016microsoft}
R.~Price, Microsoft is deleting its \uppercase{AI} chatbot’s incredibly
  racist tweets, Business Insider.

\bibitem{maguire2001methods}
M.~Maguire, Methods to support human-centred design, International journal of
  human-computer studies 55~(4) (2001) 587--634.

\bibitem{whittle2019your}
J.~Whittle, Is your software valueless?, IEEE Software 36~(3) (2019) 112--115.

\bibitem{schmidt2020interactive}
A.~Schmidt, Interactive human centered artificial intelligence: a definition
  and research challenges, in: Proceedings of the International Conference on
  Advanced Visual Interfaces, 2020, pp. 1--4.

\bibitem{amershi2014power}
S.~Amershi, M.~Cakmak, W.~B. Knox, T.~Kulesza, Power to the people: The role of
  humans in interactive machine learning, Ai Magazine 35~(4) (2014) 105--120.

\bibitem{sokol2020one}
K.~Sokol, P.~Flach, One explanation does not fit all, KI-K{\"u}nstliche
  Intelligenz 34~(2) (2020) 235--250.

\bibitem{dodge2019explaining}
J.~Dodge, Q.~V. Liao, Y.~Zhang, R.~K. Bellamy, C.~Dugan, Explaining models: an
  empirical study of how explanations impact fairness judgment, in: Proceedings
  of the 24th International Conference on Intelligent User Interfaces, 2019,
  pp. 275--285.

\bibitem{miller2017explainable}
T.~Miller, P.~Howe, L.~Sonenberg, Explainable ai: Beware of inmates running the
  asylum or: How i learnt to stop worrying and love the social and behavioural
  sciences, IJCAI 2017 Workshop on Explainable Artificial Intelligence (XAI),
  36–42, URL http://people.
  eng.unimelb.edu.au/tmiller/pubs/explanation-inmates.pdf.

\bibitem{bellamy2018ai}
R.~K. Bellamy, K.~Dey, M.~Hind, S.~C. Hoffman, S.~Houde, K.~Kannan, P.~Lohia,
  J.~Martino, S.~Mehta, A.~Mojsilovic, et~al., Ai fairness 360: An extensible
  toolkit for detecting, understanding, and mitigating unwanted algorithmic
  bias, arXiv preprint arXiv:1810.01943.

\bibitem{roselli2019managing}
D.~Roselli, J.~Matthews, N.~Talagala, Managing bias in ai, in: Companion
  Proceedings of The 2019 World Wide Web Conference, 2019, pp. 539--544.

\bibitem{dignum2017responsible}
V.~Dignum, Responsible artificial intelligence: designing ai for human values.

\bibitem{Apple2020}
{Apple Developer}, Human interface guidelines, [online]
  https://developer.apple.com/design/human-interface-guidelines/machine-learning/overview/introduction/
  [Accessed 1 May 2020].

\bibitem{GooglePair2019}
{Google Research}, The people + \uppercase{AI} guidebook, [online] Available
  at: https://research.google/teams/brain/pair/ [Accessed 1 April 2020] (2019).

\bibitem{Microsoft2022}
{Microsoft}, Guidelines for human-ai interaction, [online]
  https://www.microsoft.com/en-us/research/project/guidelines-for-human-ai-interaction/
  [Accessed 1 Feb 2022].

\bibitem{ahmad2021SLR}
K.~Ahmad, M.~Bano, M.~Abdelrazek, C.~Arora, J.~Grundy, What’s up with
  requirements engineering for artificial intelligence systems?, in: 2021 IEEE
  29th International Requirements Engineering Conference (RE), IEEE, 2021, pp.
  1--12.

\bibitem{ahmad2022mapping}
K.~Ahmad, M.~Abdelrazek, C.~Arora, M.~Bano, J.~Grundy, Requirements engineering
  for artificial intelligence systems: A systematic mapping study, arXiv
  preprint arXiv:2212.10693.

\bibitem{kuwajima2019adapting}
H.~Kuwajima, F.~Ishikawa, Adapting square for quality assessment of artificial
  intelligence systems, in: 2019 IEEE International Symposium on Software
  Reliability Engineering Workshops (ISSREW), IEEE, 2019, pp. 13--18.

\bibitem{aydemir2018roadmap}
F.~B. Aydemir, F.~Dalpiaz, A roadmap for ethics-aware software engineering, in:
  2018 IEEE/ACM International Workshop on Software Fairness (FairWare), IEEE,
  2018, pp. 15--21.

\bibitem{amaral2020ontology}
G.~Amaral, R.~Guizzardi, G.~Guizzardi, J.~Mylopoulos, Ontology-based modeling
  and analysis of trustworthiness requirements: Preliminary results, in:
  International Conference on Conceptual Modeling, Springer, 2020, pp.
  342--352.

\bibitem{hall2019systematic}
M.~Hall, D.~Harborne, R.~Tomsett, V.~Galetic, S.~Quintana-Amate, A.~Nottle,
  A.~Preece, A systematic method to understand requirements for explainable ai
  (xai) systems, in: Proceedings of the IJCAI Workshop on eXplainable
  Artificial Intelligence (XAI 2019), Macau, China, Vol.~11, 2019.

\bibitem{schoonderwoerd2021human}
T.~A. Schoonderwoerd, W.~Jorritsma, M.~A. Neerincx, K.~van~den Bosch,
  Human-centered xai: Developing design patterns for explanations of clinical
  decision support systems, International Journal of Human-Computer Studies
  (2021) 102684.

\bibitem{cirqueira2020scenario}
D.~Cirqueira, D.~Nedbal, M.~Helfert, M.~Bezbradica, Scenario-based requirements
  elicitation for user-centric explainable ai, in: International Cross-Domain
  Conference for Machine Learning and Knowledge Extraction, Springer, 2020, pp.
  321--341.

\bibitem{kohl2019explainability}
M.~A. K{\"o}hl, K.~Baum, M.~Langer, D.~Oster, T.~Speith, D.~Bohlender,
  Explainability as a non-functional requirement, in: 2019 IEEE 27th
  International Requirements Engineering Conference (RE), IEEE, 2019, pp.
  363--368.

\bibitem{MLCanvas}
{Louis Dorard}, The machine learning canvas,
  https://www.louisdorard.com/machine-learning-canvas. Accessed [March, 2020].

\bibitem{kondermann2013ground}
D.~Kondermann, Ground truth design principles: an overview, in: Proceedings of
  the International Workshop on Video and Image Ground Truth in Computer Vision
  Applications, 2013, pp. 1--4.

\bibitem{agarwal2014expert}
M.~Agarwal, S.~Goel, Expert system and it's requirement engineering process,
  in: International Conference on Recent Advances and Innovations in
  Engineering (ICRAIE-2014), IEEE, 2014, pp. 1--4.

\bibitem{shneiderman2022human}
B.~Shneiderman, Human-Centered AI, Oxford University Press, 2022.

\bibitem{ehsan2020human}
U.~Ehsan, M.~O. Riedl, Human-centered explainable ai: Towards a reflective
  sociotechnical approach, in: International Conference on Human-Computer
  Interaction, Springer, 2020, pp. 449--466.

\bibitem{xu2022transitioning}
W.~Xu, M.~J. Dainoff, L.~Ge, Z.~Gao, Transitioning to human interaction with ai
  systems: New challenges and opportunities for hci professionals to enable
  human-centered ai, International Journal of Human--Computer Interaction
  (2022) 1--25.

\bibitem{riedl2019human}
M.~O. Riedl, Human-centered artificial intelligence and machine learning, Human
  Behavior and Emerging Technologies 1~(1) (2019) 33--36.

\bibitem{wang2019designing}
D.~Wang, Q.~Yang, A.~Abdul, B.~Y. Lim, Designing theory-driven user-centric
  explainable {AI}, in: Proceedings of the 2019 CHI conference on human factors
  in computing systems, 2019, pp. 1--15.

\bibitem{hajian2016algorithmic}
S.~Hajian, F.~Bonchi, C.~Castillo, Algorithmic bias: From discrimination
  discovery to fairness-aware data mining, in: Proceedings of the 22nd ACM
  SIGKDD international conference on knowledge discovery and data mining, 2016,
  pp. 2125--2126.

\bibitem{khomh2018software}
F.~Khomh, B.~Adams, J.~Cheng, M.~Fokaefs, G.~Antoniol, Software engineering for
  machine-learning applications: The road ahead, IEEE Software 35~(5) (2018)
  81--84.

\bibitem{whittaker2019disability}
M.~Whittaker, M.~Alper, C.~L. Bennett, S.~Hendren, L.~Kaziunas, M.~Mills, M.~R.
  Morris, J.~Rankin, E.~Rogers, M.~Salas, et~al., Disability, bias, and ai, AI
  Now Institute.

\bibitem{bellamy2019Fairness360}
R.~K. Bellamy, K.~Dey, M.~Hind, S.~C. Hoffman, S.~Houde, K.~Kannan, P.~Lohia,
  J.~Martino, S.~Mehta, A.~Mojsilovi{\'c}, et~al., A{I} fairness 360: An
  extensible toolkit for detecting and mitigating algorithmic bias, IBM Journal
  of Research and Development 63~(4/5) (2019) 4--1.

\bibitem{tramer2017fairtest}
F.~Tramer, V.~Atlidakis, R.~Geambasu, D.~Hsu, J.-P. Hubaux, M.~Humbert,
  A.~Juels, H.~Lin, Fairtest: Discovering unwarranted associations in
  data-driven applications, in: 2017 IEEE European Symposium on Security and
  Privacy (EuroS\&P), IEEE, 2017, pp. 401--416.

\bibitem{Azure}
Microsoft, Azure ml, [online] Available at: https://studio.azureml.net.

\bibitem{AWS}
Amazon, Aws ml, [online] Available at:
  https://aws.amazon.com/machine-learning/.

\bibitem{khalajzadeh2018survey}
H.~Khalajzadeh, M.~Abdelrazek, J.~Grundy, J.~Hosking, Q.~He, A survey of
  current end-user data analytics tool support, in: 2018 IEEE International
  Congress on Big Data (BigData Congress), IEEE, 2018, pp. 41--48.

\bibitem{amershi2019guidelines}
S.~Amershi, D.~Weld, M.~Vorvoreanu, A.~Fourney, B.~Nushi, P.~Collisson, J.~Suh,
  S.~Iqbal, P.~N. Bennett, K.~Inkpen, et~al., Guidelines for human-{AI}
  interaction, in: Proceedings of the 2019 CHI Conference on Human Factors in
  Computing Systems, 2019, pp. 1--13.

\bibitem{villamizar2021requirements}
H.~Villamizar, T.~Escovedo, M.~Kalinowski, Requirements engineering for machine
  learning: A systematic mapping study, in: 2021 47th Euromicro Conference on
  Software Engineering and Advanced Applications (SEAA), IEEE, 2021, pp.
  29--36.

\bibitem{bruno2013functional}
B.~Bruno, F.~Mastrogiovanni, A.~Sgorbissa, Functional requirements and design
  issues for a socially assistive robot for elderly people with mild cognitive
  impairments, in: 2013 IEEE RO-MAN, IEEE, 2013, pp. 768--773.

\bibitem{sandkuhl2019putting}
K.~Sandkuhl, Putting ai into context-method support for the introduction of
  artificial intelligence into organizations, in: 2019 IEEE 21st Conference on
  Business Informatics (CBI), Vol.~1, IEEE, 2019, pp. 157--164.

\bibitem{fagbola2019towards}
T.~M. Fagbola, S.~C. Thakur, Towards the development of artificial
  intelligence-based systems: Human-centered functional requirements and open
  problems, in: 2019 International Conference on Intelligent Informatics and
  Biomedical Sciences (ICIIBMS), IEEE, 2019, pp. 200--204.

\bibitem{dimatteo2020requirements}
J.~DiMatteo, D.~M. Berry, K.~Czarnecki, Requirements for monitoring inattention
  of the responsible human in an autonomous vehicle: The recall and precision
  tradeoff., in: REFSQ Workshops, 2020.

\bibitem{cysneiros2018software}
L.~M. Cysneiros, M.~Raffi, J.~C.~S. do~Prado~Leite, Software transparency as a
  key requirement for self-driving cars, in: 2018 IEEE 26th International
  Requirements Engineering Conference (RE), IEEE, 2018, pp. 382--387.

\bibitem{rahimi2019toward}
M.~Rahimi, J.~L. Guo, S.~Kokaly, M.~Chechik, Toward requirements specification
  for machine-learned components, in: 2019 IEEE 27th International Requirements
  Engineering Conference Workshops (REW), IEEE, 2019, pp. 241--244.

\bibitem{balasubramaniam2022transparency}
N.~Balasubramaniam, M.~Kauppinen, K.~Hiekkanen, S.~Kujala, Transparency and
  explainability of ai systems: Ethical guidelines in practice, in:
  International Working Conference on Requirements Engineering: Foundation for
  Software Quality, Springer, 2022, pp. 3--18.

\bibitem{martinez2022software}
S.~Mart{\'\i}nez-Fern{\'a}ndez, J.~Bogner, X.~Franch, M.~Oriol, J.~Siebert,
  A.~Trendowicz, A.~M. Vollmer, S.~Wagner, Software engineering for ai-based
  systems: a survey, ACM Transactions on Software Engineering and Methodology
  (TOSEM) 31~(2) (2022) 1--59.

\bibitem{lu2022software}
Q.~Lu, L.~Zhu, X.~Xu, J.~Whittle, D.~Douglas, C.~Sanderson, Software
  engineering for responsible ai: An empirical study and operationalised
  patterns, in: 2022 IEEE/ACM 44th International Conference on Software
  Engineering: Software Engineering in Practice (ICSE-SEIP), IEEE, 2022, pp.
  241--242.

\bibitem{vogelsang2019requirements}
A.~Vogelsang, M.~Borg, Requirements engineering for machine learning:
  Perspectives from data scientists, IEEE International Requirements
  Engineering Conference Workshops.

\bibitem{nakamichi2020requirements}
K.~Nakamichi, K.~Ohashi, I.~Namba, R.~Yamamoto, M.~Aoyama, L.~Joeckel,
  J.~Siebert, J.~Heidrich, Requirements-driven method to determine quality
  characteristics and measurements for machine learning software and its
  evaluation, in: 2020 IEEE 28th International Requirements Engineering
  Conference (RE), IEEE, 2020, pp. 260--270.

\bibitem{berry2022requirements}
D.~M. Berry, Requirements engineering for artificial intelligence: What is a
  requirements specification for an artificial intelligence?, in: International
  Working Conference on Requirements Engineering: Foundation for Software
  Quality, Springer, 2022, pp. 19--25.

\bibitem{bosch2018takes}
J.~Bosch, H.~H. Olsson, I.~Crnkovic, It takes three to tango: Requirement,
  outcome/data, and {AI} driven development., in: SiBW, 2018, pp. 177--192.

\bibitem{abualhaija2022automated}
S.~Abualhaija, C.~Arora, A.~Sleimi, L.~C. Briand, Automated question answering
  for improved understanding of compliance requirements: A multi-document
  study, in: 2022 IEEE 30th International Requirements Engineering Conference
  (RE), IEEE, 2022, pp. 39--50.

\bibitem{challa2020faulty}
H.~Challa, N.~Niu, R.~Johnson, Faulty requirements made valuable: On the role
  of data quality in deep learning, in: 2020 IEEE Seventh International
  Workshop on Artificial Intelligence for Requirements Engineering (AIRE),
  IEEE, 2020, pp. 61--69.

\bibitem{shin2019data}
C.~Shin, S.~Rho, H.~Lee, W.~Rhee, Data requirements for applying machine
  learning to energy disaggregation, Energies 12~(9) (2019) 1696.

\bibitem{weihrauch2018conceptual}
D.~Weihrauch, P.~A. Schindler, W.~Sihn, A conceptual model for developing a
  smart process control system, Procedia CIRP 67 (2018) 386--391.

\bibitem{altarturi2017requirement}
H.~H. Altarturi, K.-Y. Ng, M.~I.~H. Ninggal, A.~S.~A. Nazri, A.~A. Abd~Ghani, A
  requirement engineering model for big data software, in: 2017 IEEE Conference
  on Big Data and Analytics (ICBDA), IEEE, 2017, pp. 111--117.

\bibitem{ries2021mde}
B.~Ries, N.~Guelfi, B.~Jahic, An {MDE} method for improving deep learning
  dataset requirements engineering using alloy and {UML}, in: Proceedings of
  the 9th International Conference on Model-Driven Engineering and Software
  Development, SCITEPRESS, 2021, pp. 41--52.

\bibitem{horkoff2019nonFunctional}
J.~Horkoff, Non-functional requirements for machine learning: Challenges and
  new directions, in: 2019 IEEE 27th International Requirements Engineering
  Conference (RE), IEEE, 2019, pp. 386--391.

\bibitem{habibullah2021non}
K.~M. Habibullah, J.~Horkoff, Non-functional requirements for machine learning:
  Understanding current use and challenges in industry, IEEE 29th International
  Requirements Engineering Conference (RE).

\bibitem{cysneiros2020non}
L.~M. Cysneiros, J.~C.~S. do~Prado~Leite, Non-functional requirements orienting
  the development of socially responsible software, in: Enterprise,
  Business-Process and Information Systems Modeling, Springer, 2020, pp.
  335--342.

\bibitem{sculley2015hidden}
D.~Sculley, G.~Holt, D.~Golovin, E.~Davydov, T.~Phillips, D.~Ebner,
  V.~Chaudhary, M.~Young, J.-F. Crespo, D.~Dennison, Hidden technical debt in
  machine learning systems, in: Advances in neural information processing
  systems, 2015, pp. 2503--2511.

\bibitem{krause2016interacting}
J.~Krause, A.~Perer, K.~Ng, Interacting with predictions: Visual inspection of
  black-box machine learning models, in: Proceedings of the 2016 CHI Conference
  on Human Factors in Computing Systems, 2016, pp. 5686--5697.

\bibitem{bonfe2012towards}
M.~Bonfe, F.~Boriero, R.~Dodi, P.~Fiorini, A.~Morandi, R.~Muradore,
  L.~Pasquale, A.~Sanna, C.~Secchi, Towards automated surgical robotics: A
  requirements engineering approach, in: 2012 4th IEEE RAS \& EMBS
  International Conference on Biomedical Robotics and Biomechatronics (BioRob),
  IEEE, 2012, pp. 56--61.

\bibitem{kitchenham2002design}
B.~A. Kitchenham, S.~L. Pfleeger, Principles of survey research part 2:
  designing a survey, ACM SIGSOFT Software Engineering Notes 27~(1) (2002)
  18--20.

\bibitem{Jira}
J.~Software, Atlassian solution, [online] Available at:
  https://www.atlassian.com/software/jira [Accessed 20 December 2021].

\bibitem{Confluence}
C.~Software, Atlassian solution, [online] Available at:
  https://www.atlassian.com/software/confluence [Accessed 20 December 2021].

\bibitem{SPSS}
S.~Software, Ibm solutions, [online] Available at:
  https://www.ibm.com/au-en/products/spss-statistics [Accessed 20 December
  2021].

\bibitem{DOORS}
D.~software, Ibm solutions, [online] Available at:
  https://www.ibm.com/au-en/products/requirements-management [Accessed 20
  December 2021].

\bibitem{libreoffice}
libreoffice, [online] Available at: https://www.libreoffice.org [Accessed 20
  December 2021].

\bibitem{EA}
E.~Architect, Sparx systems, [online] Available at:
  https://www.sparxsystems.com [Accessed 20 December 2021].

\bibitem{Pencil}
P.~Software, [online] Available at: https://pencil.evolus.vn [Accessed 20
  December 2021].

\bibitem{Drawio}
D.~Software, [online] Available at: https://drawio-app.com [Accessed 20
  December 2021].

\bibitem{neace2018goal}
K.~Neace, R.~Roncace, P.~Fomin, Goal model analysis of autonomy requirements
  for unmanned aircraft systems, Requirements Engineering 23~(4) (2018)
  509--555.

\bibitem{dimitrakopoulos2019alpha}
G.~Dimitrakopoulos, E.~Kavakli, P.~Loucopoulos, D.~Anagnostopoulos,
  T.~Zographos, A capability-oriented modelling and simulation approach for
  autonomous vehicle management, Simulation Modelling Practice and Theory 91
  (2019) 28--47.

\bibitem{amyot2011grl}
D.~Amyot, G.~Mussbacher, S.~Ghanavati, J.~Kealey, {GRL} modeling and analysis
  with {jUCMNav}, iStar 766 (2011) 160--162.

\bibitem{viyovic2014sirius}
V.~Viyovi{\'c}, M.~Maksimovi{\'c}, B.~Perisi{\'c}, Sirius: A rapid development
  of dsm graphical editor, in: IEEE 18th International Conference on
  Intelligent Engineering Systems INES 2014, IEEE, 2014, pp. 233--238.

\bibitem{herrmann2013requirements}
A.~Herrmann, Requirements engineering in practice: There is no requirements
  engineer position, in: International Working Conference on Requirements
  Engineering: Foundation for Software Quality, Springer, 2013, pp. 347--361.

\bibitem{wang2018understanding}
C.~Wang, P.~Cui, M.~Daneva, M.~Kassab, Understanding what industry wants from
  requirements engineers: An exploration of re jobs in canada, in: Proceedings
  of the 12th ACM/IEEE International Symposium on Empirical Software
  Engineering and Measurement, 2018, pp. 1--10.

\bibitem{daneva2017job}
M.~Daneva, C.~Wang, P.~Hoener, What the job market wants from requirements
  engineers? an empirical analysis of online job ads from the netherlands, in:
  2017 ACM/IEEE international symposium on empirical software engineering and
  measurement (ESEM), IEEE, 2017, pp. 448--453.

\end{thebibliography}


\end{document}